\documentclass{iopjournal}
\expandafter\let\csname equation*\endcsname\relax
\expandafter\let\csname endequation*\endcsname\relax

\usepackage{amsmath}
\usepackage{amssymb}  
\usepackage{mathtools} 
\usepackage{braket}
\usepackage{bbm} 
\usepackage{graphicx}
\usepackage{xcolor}
\usepackage{ulem}  
\usepackage{todonotes}
\usepackage{subcaption}
\usepackage{placeins} 

\DeclareMathOperator{\hc}{\mathrm{h.c.}}
\DeclareMathOperator{\elliptic}{\mathrm{E}}

\newcommand{\ketbra}[2]{\ket{#1}\bra{#2}}

\newcommand{\rfac}{r_\mathrm{f}}
\newcommand{\potnill}{V_0}
\newcommand{\vnn}{\potnill}

\newcommand{\rosc}{r_\mathrm{osc}}

\newcommand{\omegatildeq}{\widetilde{\omega}_q}
\newcommand{\omegatilde}{\widetilde{\omega}}
\newcommand{\hphon}{\hat{h}_\mathrm{p}}

\newcommand{\ktwo}{\kappatwo}

\newcommand{\kappaone}{\kappa^{(1)}}

\newcommand{\kappatwo}{\kappa^{(2)}}

\newcommand{\costerm}{\cos\frac{\pi q}{m}}
\newcommand{\Cos}[1]{\cos \left( #1 \right)}

\newcommand{\opsx}{\hat{\sigma}^x}
\newcommand{\opn}{\hat{n}}
\newcommand{\opa}{\hat{a}}
\newcommand{\opadag}{\hat{a}^\dagger}

\newcommand{\opcapitala}{\hat A}
\newcommand{\opcapitaladag}{\hat A^\dagger}
\newcommand{\opcapitaladagsq}{\hat A^{\dagger 2}}
\newcommand{\opcapitalasq}{\hat A^2}

\newcommand{\opd}{\hat{d}}
\newcommand{\opddag}{\hat{d}^\dagger}

\newcommand{\opg}{\hat{g}}
\newcommand{\opgdag}{\hat{g}^\dagger}

\newcommand{\spinup}{\ket{\uparrow}}
\newcommand{\spindn}{\ket{\downarrow}}
\newcommand{\braspinup}{\bra{\uparrow}}
\newcommand{\braspindn}{\bra{\downarrow}}

\begin{document}


\title{Quantum-Noise Induced Localization and Motional Squeezing in a  Rydberg Quantum
Simulator}

\author{Benno Bock$^{1,*+}$\orcid{0009-0000-8613-8198}, Daniel Brady$^{1,+}$\orcid{0000-0003-3993-7098} and Michael Fleischhauer$^{1}$\orcid{0000-0003-4059-7289}}

\affil{$^1$Department of Physics and Research Center OPTIMAS, RPTU University Kaiserslautern-Landau, D-67663 Kaiserslautern, Germany}

\affil{$^*$corresponding author}

\affil{$^+$these authors contributed equally}

\email{bbock@rptu.de}


\begin{abstract}
We investigate the interplay between mechanical forces and the internal-state dynamics 
of Rydberg excitations in atom-tweezer arrays. Dipole interactions between Rydberg atoms facilitate excitation spreading, but at the same time couple electronic (spin) degrees of freedom with motional (phonon) states.  With increasing spin-phonon coupling, the growth dynamics of a cluster of excited Rydberg atoms changes from ballistic spreading to Bloch-like oscillations and eventually to Anderson-like localization.
We show that these effects are caused by quantum fluctuations in the phonon field: The dynamics of a Rydberg cluster can be mapped to a single particle in a semi-infinite lattice subject to phonon-induced energy shifts. The mean-field contribution of this energy shift leads to a linear potential gradient, resulting into Bloch-like oscillations. In addition,  quantum fluctuations of phonons create a random local potential causing a transition from a regime of Bloch oscillations to localization.
The spin-phonon coupling leads furthermore
to highly correlated and non-classical phonon states in the form of squeezed states of the position of the Rydberg atoms. Depending on the form of the dipolar interaction potential, either in- or out-of-phase correlated oscillations of atoms emerge.
\end{abstract}

\section{Introduction}

Rydberg atoms have become a powerful tool for constructing 
neutral-atom quantum simulators and quantum information systems \cite{browaeys2020many} due to their strong, long-range dipole interactions. These interactions are typically on the order of GHz and on µm length scales \cite{gallagher1994rydberg} and can be tailored simply by adjusting laser parameters.
With advances in ultra-cold atom trapping using tweezer arrays, arbitrary geometries of neutral atoms can be programmed \cite{barredo2016atom, endres2016atom, barredo2018synthetic}. Through this high level of experimental control, Rydberg simulators of many-body quantum spin systems \cite{weimer2010rydberg} have found a wide use of applications, for example to study the quantum Ising model \cite{weimer2008quantum, labuhn2016tunable, schauss2015crystallization, lienhard2018observing, guardado2018probing}, coherent transport properties \cite{barredo2015coherent}, modeling topological systems 
\cite{de2019observation}, or quantum phase transitions to $\mathbb{Z}_\mathrm{N}$ symmetric phases 
\cite{schachenmayer2010dynamical, bernien2017probing, samajdar2018numerical}, and spin liquid phases \cite{Semeghini2021, ohler2023quantum}.

\begin{figure}
    \centering
    \begin{subfigure}{0.45\textwidth}
    \includegraphics[width=\linewidth]{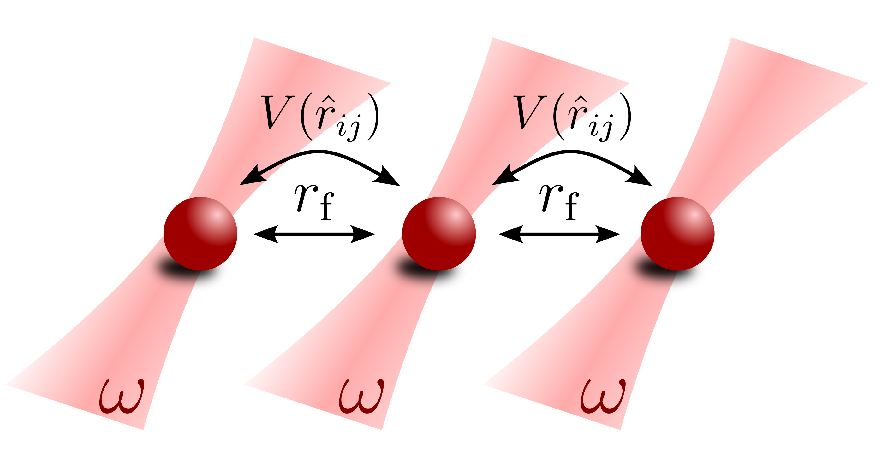}  
    \caption{}
    \end{subfigure}
    \hfill
    \begin{subfigure}{0.45\textwidth}
    \includegraphics[width=\linewidth]{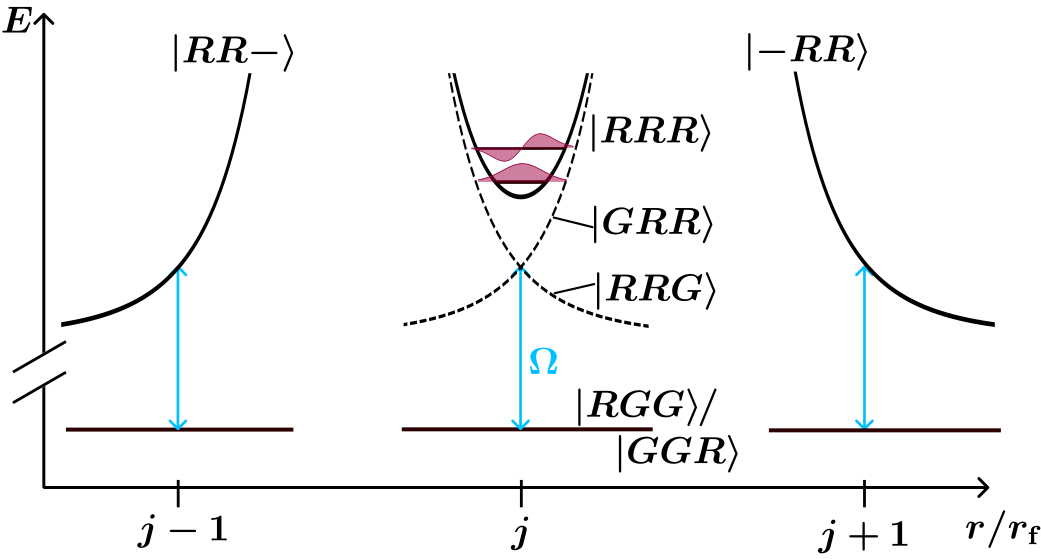}    
    \caption{}
    \end{subfigure}
    \caption{(a) Representation of interacting atoms in tweezer traps. (b) Van-der-Waals potential and the resulting potential for atoms in the center of a Rydberg atom domain. In the bulk of an excitation domain dipole forces from adjacent Rydberg atoms compensate and lead to a harmonic confinement potential.}
     \label{fig:hphon}
\end{figure}

A key problem of Rydberg quantum simulators are the mechanical forces accompanying the dipole-dipole interactions, which can lead to a dephasing of optical transitions \cite{li2013nonadiabatic}
and mechanical instabilities. Consequently, quantum simulations involving Rydberg blockade \cite{jaksch2000fast,lukin2001dipole} or Rydberg facilitation \cite{ates2007antiblockade} 
are typically performed on short time scales, where motional effects can be neglected.
Here we show for the example of a quantum contact process based on Rydberg facilitation (see Fig.~\ref{fig:hphon}) that mechanical forces give rise to very rich physics. They can strongly modify the dynamics of excitations and induce non-classical motional states as well as spin-phonon correlations.

Specifically we investigate 
the dynamics of interacting Rydberg atoms in a 1D tweezer array in the facilitation regime, representing a quantum contact process.
Dipole interactions couple the electronic (spin) degrees of freedom with excited motional states (phonons) in the tweezer traps. 
For very weak spin-phonon couplings the excitation spreading is ballistic. When increasing the coupling, the mean-field energy shift due to phonons leads
to an oscillatory dynamics in the Rydberg facilitation akin of Bloch oscillations in a lattice. For even stronger couplings the spreading stops and the distribution of excitations approaches a localized state, see Fig.~\ref{fig:rydberg-density-plot}.

\section{Model: Rydberg Quantum Simulator}

We consider a 1D chain of $N$ atoms in tweezer traps with lattice spacing $\rfac$ each having an internal electronic (spin) and a vibrational (phonon) degree of freedom with trapping frequency $\omega$, see Fig.~\ref{fig:hphon} a). The atoms are laser coupled with a Rabi frequency $\Omega$ between a ground~$\spindn$ and a high lying Rydberg~$\spinup$ state. Furthermore, the laser has a large detuning from resonance $\Delta$, such that ${\Delta\gg\Omega}$. Strong dipolar interactions between Rydberg atoms give rise to a Rydberg-Rydberg interaction potential $V(\hat{r}_{ij})$, where $\hat{r}_{ij}$ corresponds to the interatomic distance. 
The full Hamiltonian is given by
\begin{equation}
    \label{eq:base_hamiltonian}
    \hat{H} = \sum_{j=1}^{N}
    \Omega \opsx_j + \Delta \opn_j + \omega 
    \Big( 
        \opadag_j \opa_j + \frac{1}{2}
    \Big)
    + V(\hat{r}_{j,j+1}) \opn_j \opn_{j+1},
\end{equation}
with 

${\opsx = \spinup \braspindn + \spindn \braspinup}$, and projection operator onto Rydberg states ${\opn = \ket{\uparrow} \bra{\uparrow}}$, and ${\hbar = 1}$. We also restrict the range of interactions to nearest neighbors.
Typically the dipole potential takes the form of a van-der-Waals (vdW) potential, i.e. ${V(\hat{r}_{ij}) \sim \hat{r}_{ij}^{-6}}$ \cite{gallagher1994rydberg}, which is nearly linear at the lattice spacing $\rfac$, and gives rise to strong repulsive (or attractive) forces between atoms in the Rydberg state.
We therefore expand the potential around the lattice spacing $r_f$ up to second order in the interatomic displacement:
\begin{align}
    \nonumber
    V(\hat r_{j,j+1}) = 
    \potnill
    + 
    V^\prime (\rfac) (\hat r_{j,j+1} - \rfac)
    +
    \frac{1}{2} V^{\prime \prime}(\rfac)
    (\hat r_{j,j+1} - \rfac)^2
    + \mathcal{O}(\hat r_{j,j+1}^3),
\end{align}
with ${\potnill \equiv V(\rfac)}$.
The relative distance can be expressed as ${\hat{r}_{j, j+1} = \rfac + \hat{x}_{j+1} - \hat{x}_j}$, where $\hat{x}_j$ is the position operator of the $j$th atom relative to the center of tweezer trap $j$. Expressing the position operator in terms of bosonic creation and annihilation operators, i.e. ${\hat{x}_j = \rosc (\opadag_j + \opa_j)}$ with $\rosc = \sqrt{\frac{1}{2m\omega}}$, we receive
\begin{align}
    V(\hat r_{j,j+1}) \approx
    \potnill 
    \Big[
    1 - 6\kappa 
    \big(
    \hat a^\dagger_{j+1} - \hat a^\dagger_j + \mathrm{h.c.}
    \big)
    + 21\kappa^2 
    \big(
    \hat S_j + \hat S_{j+1} - 2 \hat T_{j,j+1}
    \big)
    \Big].
    \label{eq:approximated-pot}
\end{align}
Here we introduced the operators ${\hat S_j=\hat{a}^{\dagger 2}_j + \opa_j^2 + 2 \opadag_j \opa_j + 1}$ describing \textit{local} squeezing terms and ${\hat T_{j, j+1} = \opadag_{j+1} \opadag_j + \opadag_{j+1} \opa_j + \mathrm{h.c.}}$ describing \textit{non-local} pair-creation/annihilation, as well as phonon transport terms.
The parameter $\kappa = \frac{r_\mathrm{osc}}{\rfac}$ corresponds to the ratio of trap width to tweezer spacing.
We in addition introduce the parameters $\kappaone = 6\vnn\kappa$ and $\kappatwo = 21\vnn\kappa^2$ representing the energy scales of the respective derivative of the potential.  If the value of $\vnn$ is fixed, one of these parameters is determined by the other.
\\
Finally, we consider the system under the \textit{facilitation constraint}, i.e. ${\vnn + \Delta = 0}$, where the detuning cancels out the dipole potential at the lattice spacing $\rfac$. As a result, atoms neighbored by exactly \textit{one} Rydberg atom are resonantly coupled to the light field.

\section{Dynamics of Spin Domains}
\begin{figure}
    \centering
    \includegraphics[width=\linewidth]{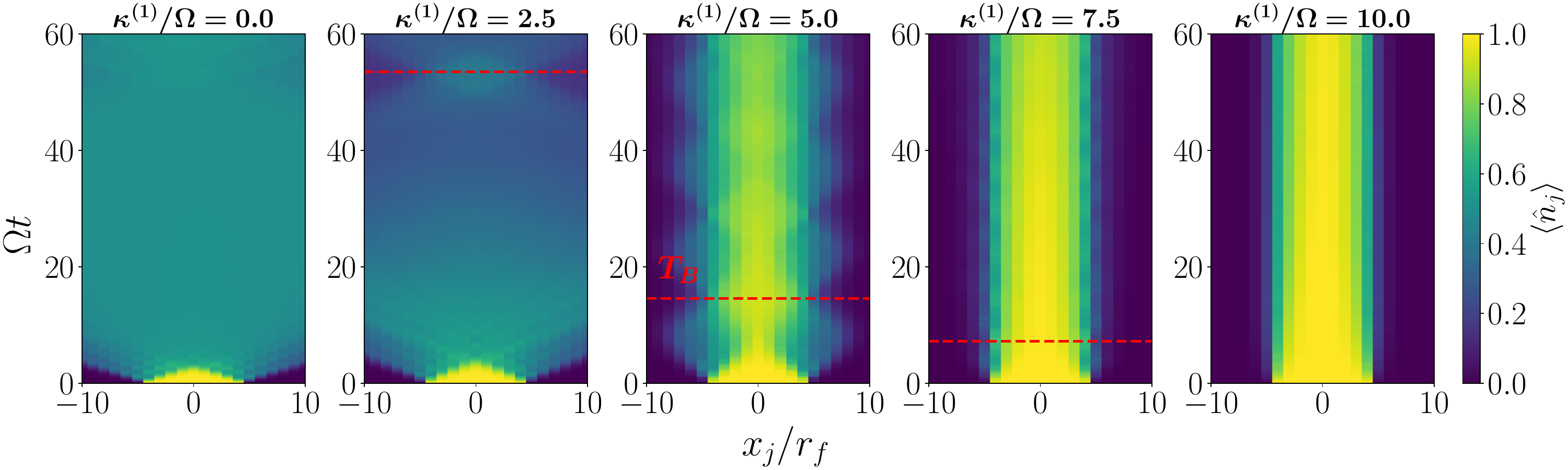}
    \caption{Numerical results: For increasing spin-phonon coupling, the spin domain goes from expanding ballistically, to oscillating to localizing. Red lines represent the analytically calculated oscillation periods Eq.~\eqref{eq:bloch_period}, which agree perfectly with the numerical results.} 
    \label{fig:rydberg-density-plot}
\end{figure}

First, we want to investigate the dynamics of spin domains.
We perform numerical simulations using a time evolving block decimation (TEBD) algorithm \cite{vidal2003efficient} on the Hamiltonian~\eqref{eq:base_hamiltonian} with the potential given in Eq.~\eqref{eq:approximated-pot}. 
For all simulations in this paragraph we use ${\omega = 5 \Omega}$, ${N=100}$, ${\potnill = 60 \Omega}$, and we truncate the local phonon Hilbert space at $n_\mathrm{max} = 7$.
We tune $\kappa$ such that $0\leq\kappaone\lesssim10$.
We then simulate the time evolution of a single spin domain of size $m=9$, where all atoms including the Rydberg atoms are initially in their motional ground state, i.e. $\opadag_j\opa_j = 0$.
The results are shown in Fig.~\ref{fig:rydberg-density-plot} . 
It can be observed that the behavior of the system varies drastically depending on the strength $\kappaone$ of the spin-phonon coupling. 
For $\kappaone = 0$, the spin domain is expanding ballistically in agreement with results of Ref.\cite{magoni2024coherent}. For intermediate strengths of $\kappaone /\Omega\approx 5$, the spin domain size oscillates in time, while for strong interactions, the dynamics come to a complete halt, resulting in a stationary spin domain with smoothed out edges.

\subsection{Effective Single-Particle Model}

\begin{figure}[h]
    \centering
    \includegraphics[width=0.7\linewidth]{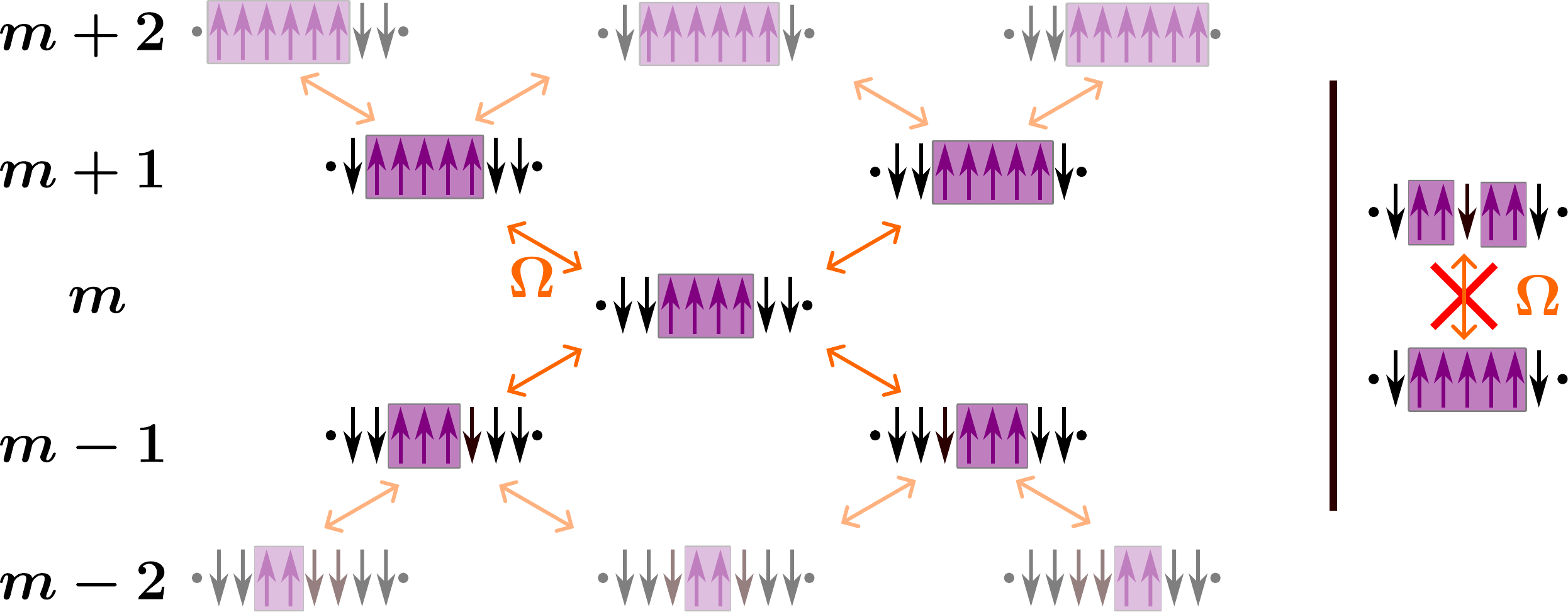}
    \caption{Symmetric coupling of spin domains. Since the Hamiltonian conserves parity, the Rabi driving does not couple individual spin domains, but rather symmetric superpositions of domains, so that the center of mass is conserved on average. This leads to a ``symmetric coupling ladder'', where spin domain states of different sizes are now characterized only by their size, depending on the initial domain (left).
    Because of the facilitation constraint, a spin domain cannot resonantly be split, nor can two spin domains coalesce (right).
    }
    \label{fig:symmmetric-coupling}
\end{figure}

As a result of the strong detuning $\Delta$, only atoms with a single Rydberg neighbor are resonantly laser coupled. 
Hence the many-body spin dynamics reduce to the dynamics of spin domains which can either grow or shrink at the edges with rate $\Omega$. However, due to Rydberg blockade two domains cannot coalesce or split \cite{magoni2021emergent}(see Fig.~\ref{fig:symmmetric-coupling}).

As pointed out in Ref.~\cite{magoni2021emergent}, a spin domain is characterized by its size $m$ and center of mass $c$, i.e. a chain of 4 Rydberg atoms at sites $j = 0,1,2,3$ can be expressed as $\ket{\Psi} = \ket{m=4, c=\frac{3}{2}}$, where $c$ is always (half-) integer if $m$ is odd (even).
In order to find out the action of the Hamiltonian Eq.~\eqref{eq:base_hamiltonian} on such a state, we divide it two parts $\hat{H} = \hat{H}_x + \hat{H}_z$, where $\hat{H}_x = \sum_{j=1}^\infty \Omega\hat\sigma_j^x$ and $\hat{H}_z$ is the rest.
We note that only $\hat{H}_x$ can change a spin domain state, since $\Psi$ is constructed of eigenstates of $\hat \sigma ^z$ and therefore $\hat{H}_z$.
Because of the facilitation constraint, only spins at the edge of a spin domain can be flipped.
Furthermore, the Hamiltonian conserves parity.
An initial spin domain state $\ket{m,c}$ is therefore resonantly coupled to the symmetric superposition of flipping the right and left spin of the spin domain $\ket{m\pm1, c}_S$, by which we mean 
\begin{equation}
    \ket{m\pm1, c}_S \coloneq \frac{1}{\sqrt{2}}\bigl(\ket{m \pm 1, c-\frac{1}{2}} + \ket{m\pm1, c+\frac{1}{2}}\bigr).
\end{equation}
The action of $\hat{H}_x$ on such a state can therefore shortly be written as
\begin{equation}
    \hat{H_x}\ket{m, c} = \Omega\bigl(\ket{m+1, c}_S + \ket{m-1, c}_S\bigr).
\end{equation}
Starting from an initial spin domain state, i.e. $\ket{m_0, c=0}\coloneqq \ket{m_0}$, we can generalize this result to construct any symmetric superposition state by going up and down this ``coupling ladder'' (see Fig.~\ref{fig:symmmetric-coupling} for a visualization)
\begin{equation}
    \ket{m_0\pm n} \coloneq  \frac{1}{\sqrt{\mathcal{N}_n}}\sum_{k=0}^n\binom{n}{k}\ket{m_0\pm n, -\frac{n}{2} + k}.
    \label{eq:general_superposition_state}
\end{equation}
with
\begin{equation}
    \mathcal{N}_n = \sum_{k=0}^n\binom{n}{k} \equiv 2^n.
\end{equation}
By induction, it then follows that the action of $\hat{H}_x$ leaves this subspace invariant. For better clarity we only consider the case $m\to m+1$, i.e.  the action of $\hat H_+ = \Omega \sum_j \hat \sigma_j^+$:
\begin{equation}
\begin{aligned}
\hat H_+ \ket{m_0\pm n} 
&= \frac{1}{\sqrt{2^n}} \sum_{k=0}^n \binom{n}{k} \hat H_+ \ket{m_0\pm n ,\,-\tfrac{n}{2}+k} \\[4pt]
&= \frac{\Omega}{\sqrt{2^{n+1}}} \sum_{k=0}^n \binom{n}{k} 
   \Bigl( \ket{m_0\pm n+1,\,-\tfrac{n}{2}+k-\tfrac12} 
        + \ket{m_0\pm n+1,\,-\tfrac{n}{2}+k+\tfrac12} \Bigr) \\[4pt]
&= \frac{\Omega}{\sqrt{2^{n+1}}} \left[
   \sum_{k=0}^n \binom{n}{k} \Bigl(\ket{m_0\pm n+1,\,-\tfrac{n+1}{2}+k} + \ket{m_0\pm n+1,\,-\tfrac{n+1}{2}+k+1}\Bigr) \right] \\[4pt]
&= \frac{\Omega}{\sqrt{2^{n+1}}} \left[
\binom{n}{0} + \sum_{k=1}^n \left(\binom{n}{k}+\binom{n}{k-1}\right)+\binom{n}{n}
\right]\ket{m_0\pm n+1,\,-\tfrac{n+1}{2}+k} \\[4pt]
&= \frac{\Omega}{\sqrt{2^{n+1}}} \sum_{k=0}^{n+1} \binom{n+1}{k} \ket{m_0\pm n+1,\,-\tfrac{n+1}{2}+k} \\[4pt]
&\equiv \Omega \ket{m_0\pm n+1}
\end{aligned}
\end{equation}
Therefore, the dynamics of the spin domain reduces to a ladder, with individual states being characterized only by their size $m$ and given by Eq.~\eqref{eq:general_superposition_state}.
Utilizing this, we can describe the spin domain of size $m$ as a particle at position $m$ in a semi-infinite lattice \cite{magoni2021emergent} with hopping amplitude $\Omega$, and the Hamiltonian reduces to
\begin{align}
    \nonumber
    \hat H = \sum_{m=1}^\infty
    &\Omega (\ket{m} \bra{m+1} + \mathrm{h.c.})
    \\
    &+ 
    \label{eq:tb_hamiltonian}
    \Big[
        m\Delta + (m-1) \potnill + \hat{h}_p(m)
    \Big]
    \ket{m} \bra{m}.
\end{align}
Furthermore, under the facilitation constraint, i.e. ${\Delta + \potnill = 0}$, there is a site dependent energy given by the total phonon energy of all atoms in the domain, i.e.
\begin{equation}
    \hphon(m) 
= \sum_{l=0}^{m-1}\omega (\opadag_l \opa_l + \tfrac{1}{2})
  + \sum_{l=0}^{m-2}\bigl(\hat{{V}}_{l,l+1} - \vnn\bigr).
\end{equation}
This Hamiltonian is diagonalized in the Supplemental Material using Bogoliubov transformed phononic operators $\opg_q$, and reads
\begin{equation}
    \hphon(m) = \sum_{q=0}^{m-1}\left[\omegatildeq\bigl(\opgdag_q\opg_q + \frac{1}{2}\bigr)\right] + \delta(m),
\end{equation}
with the dispersion relation
\begin{equation}
    \omegatildeq = \sqrt{\omega^2 + 8\omega\kappatwo\Bigl(1-\costerm\Bigr)}.
\end{equation}
$\delta(m)$ is a c-number contribution depending on the domain length and is given explicitly in the Supplementary Material.

For sufficiently large cluster sizes $m$, where the continuum approximation is valid for  $k = \frac{\pi q}{m}$, 
$\hat g^\dagger(k) \hat g(k)$ is a constant of motion, and only the vacuum term in the phonon Hamiltonian $\hat h_p(m)$ depends on the cluster size $m$.

\subsection{Wannier-Stark Regime}
%
In mean-field approximation we treat the sum of the number operators of Bogoliubov phonons as a (constant) c-number $C$ and the Hamiltonian reduces to:
\begin{equation}
    \hat H_\textrm{MF} = \sum_{m=1}^\infty \Omega \Bigl(\ket{m} \bra{m+1} + \mathrm{h.c.}\Bigr) +  \epsilon_0(m) \, \vert m\rangle\langle m\vert,
\end{equation}
where 
\begin{equation}
    \epsilon_0(m) =   C -\potnill +  m\Big[\Delta +\potnill + \frac{\braket{\tilde \omega(k)}_k}{2} - \frac{\omega}{2}\Big]
\end{equation}
and $m\,\frac{\braket{\tilde \omega(k)}_k}{2}$ is the average vacuum energy of a $k$-mode in the $\opg$-basis.

Importantly the on-site energy $\epsilon_0(m)$ is \textit{linear} in $m$, which corresponds to a potential gradient. Such a system is known to show Bloch oscillations \cite{bloch1929quantenmechanik}. The Bloch period is given by ${T_B = 2 \pi / |\partial_m \epsilon_0|}$. Under the facilitation constraint, $\Delta+\potnill=0$, ${T_B}$ only depends on the difference in phonon ground state energy between the $\opa$ and $\opg$ basis, and is given by
\begin{align}
    \label{eq:bloch_period}
    T_B = \frac{4\pi}{|\braket{\tilde \omega(k)}_k - \omega|}.
\end{align}
In Fig.~\ref{fig:rydberg-density-plot} we have shown the cluster dynamics starting from an initial cluster of size $m_0=9$, obtained from TEBD simulations of the microscopic Hamiltonian~\eqref{eq:base_hamiltonian} with 
\eqref{eq:approximated-pot} for different values of the spin-phonon coupling strength $\kappa$. The density plots with visible Bloch oscillations show periods that agree
perfectly with Eq.~\eqref{eq:bloch_period}.

%
\subsection{Localization Regime}
\begin{figure}
    \centering
    \includegraphics[width=\linewidth]{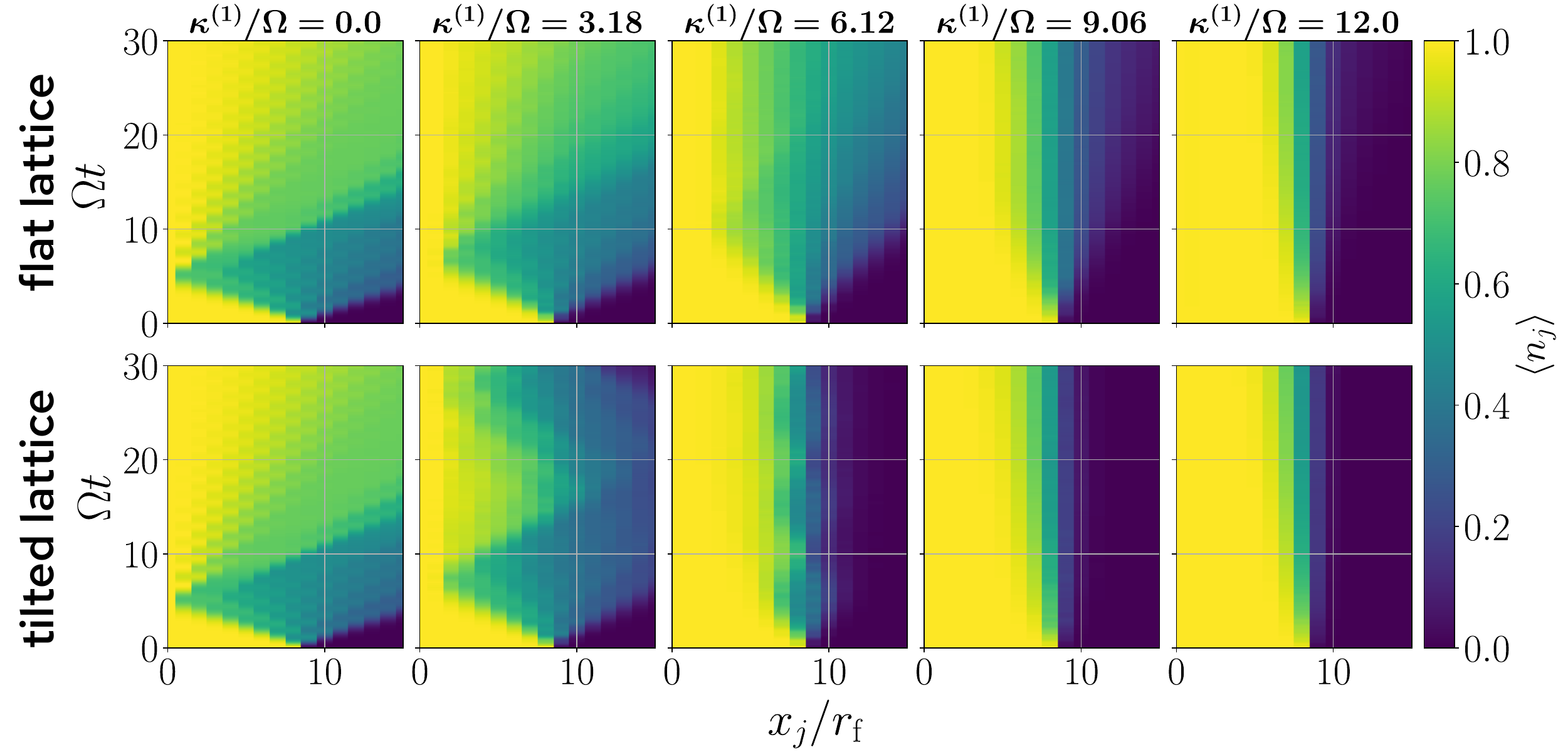}
    \caption{Rydberg density without slope in the effective Hamiltonian ($\partial_m\epsilon_0 = 0$). One can see that for a strong spin phonon coupling strength, the spin domain localizes even for a flat super lattice, indicating that fluctuations of $\hphon(m)$ cause an effective disorder and thereby localization.}
    \label{fig:flat-lattice}
\end{figure}

%
Finally, we want to investigate the behavior of the spin domain for strong spin-phonon coupling, which requires to go beyond the mean-field approximation of the phonon Hamiltonian.
To this end, we note the following: As shown in the Supplementary data, the initial state of the system, i.e. vacuum of the $\opa$-phonons, $\vert 0\rangle$ is not an eigenstate of the phonon number $\hat n_g(k) = \hat g^\dagger(k)\hat g(k)$ of the Bogoliubov modes, but has a probability distribution
\begin{equation}
    p_k(n) = \bigl\vert \langle 0\vert n\rangle_k\bigr\vert^2,\qquad \hat n_g \vert n\rangle = n \vert n\rangle, \qquad n \in \mathbb{N}.
\end{equation}
Consequently the local energy $\epsilon(m)$ of the effective Hamiltonian is random with quenched and - as we will see later - correlated disorder.
Such a system is known to show Anderson Localization \cite{Anderson1958}.
\\
In order exclude the influence of other factors, such as the simultaneously decreasing oscillation amplitude, we choose a detuning that flattens the lattice in the effective model, i.e. we place the system under the modified facilitation constraint:
\begin{equation}
    \Delta + \potnill + \frac{\langle\widetilde{\omega}(k)\rangle_k - \omega}{2} = 0.
    \label{eq:flat-lattice-condition}
\end{equation}
Then the only contribution to the on-site energy comes from the fluctuations of the phonon Hamiltonian $\delta\hphon(m)$
\begin{equation}
        \hat H = \sum_{m=1}^\infty \Omega \bigl(\ket{m} \bra{m+1} + \mathrm{h.c.}\bigr) +  \delta \hphon(m)  \ketbra{m}{m}.
\end{equation}
Additionally, in order to be able to map the spin domain states to the single particle wavefunction $\psi(m)$, we simulate a slightly modified system, where the excitation can only spread to one side starting from an initial, non-driven Rydberg atom.
This unidirectional model shows the same features as the symmetrically expanding one. However, mapping the Rydberg state to the single particle wavefunction with \textit{local} potential $\delta \hat h_p(m)$ is  only possible in this model, where an effective single particle basis state is given by
\begin{equation}
    \ket{m} = \prod_{j = 1}^{m} \ket{\uparrow}_j\prod_{j = m+1}^{N} \ket{\downarrow}_j
\end{equation}
and not the symmetric superposition of many configurations as in eq.~\eqref{eq:general_superposition_state}. Hence
\begin{equation}
    \psi(m) = \braket{m|\Psi}
    \label{eq:m-projector}
\end{equation}
where $\ket{\Psi}$ is the full numerical state.
The numerical simulations of this uni-directional spreading model with the tilted and the artificially flattened lattice is shown in Fig.~\ref{fig:flat-lattice}.
Without quantum fluctuations, we would expect a ballistic expansion of the spin domain in the whole parameter regime $0\leq\kappaone / \Omega\leq12$ in the flat lattice.
As can be seen however, the spin domain localizes for large coupling strengths even in the flat lattice, where quantum fluctuations are the only contribution to the local potential, indicating that indeed, quantum fluctuations of $\hphon$ are responsible for the localization.
From these simulations, we then compute the localization length by projecting the numerical state to domain sizes via eq.~\eqref{eq:m-projector} and fitting the wavefunction of the last time step to an exponential $\psi_\mathrm{fit}(m) = \mathcal{N}\;e^{-\frac{|m-m_0|}{\Lambda}}$, where $m_0$ is the initial domain size.

\begin{figure}
    \centering
    \includegraphics[width = \linewidth]{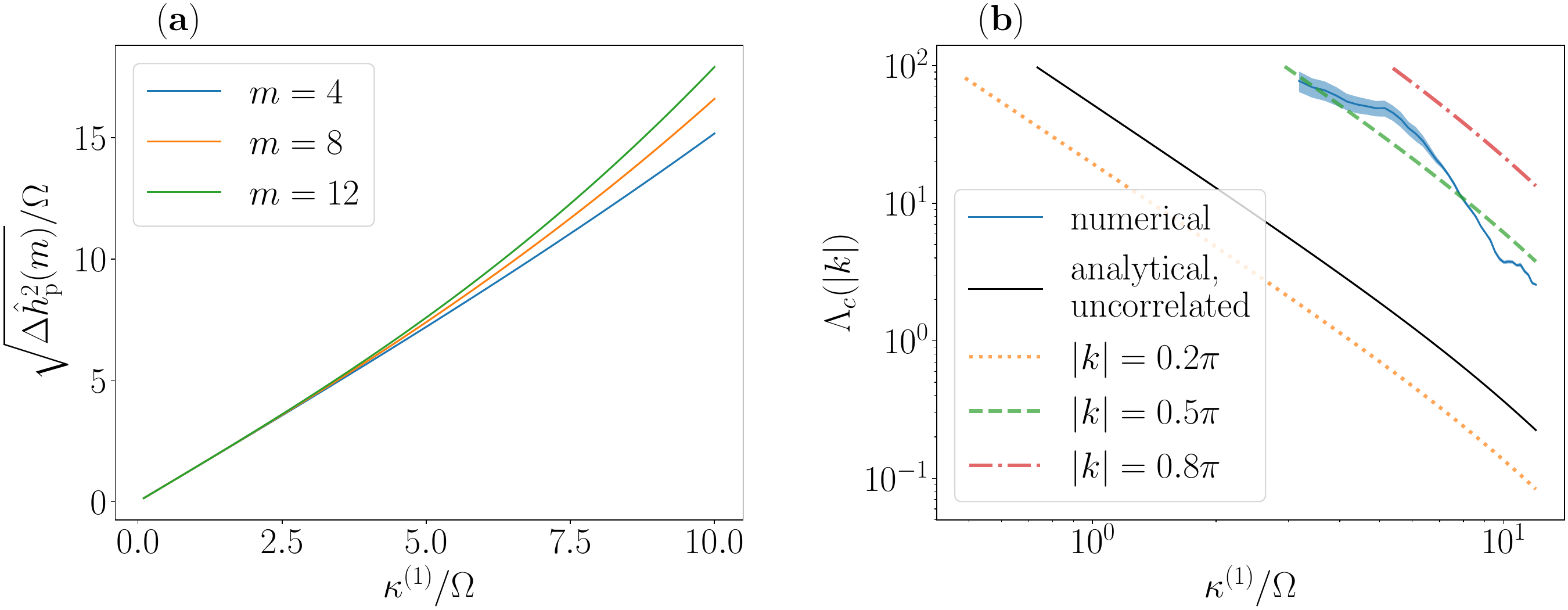}
    \caption{ (a) Standard deviation of the Phonon Hamiltonian calculated in the Supplementary Material. (b)Localization length as a function of $\kappaone$. One observes highest agreement between numerical results (blue line) if correlation is included and for a momentum $|k| = \frac{\pi}{2}$.}
    \label{fig:loc-length}
\end{figure}

In order to quantify the localization, we calculate the localization length by assuming the disorder to be uniformly distributed around the mean value $\braket{\hphon(m)}$ with the width being the standard deviation of the phonon Hamiltonian $W = \sqrt{\Delta \hphon(m)^2}$, which we calculate in the Supplementary Material.
For random disorder, the localization length $\Lambda$ reads \cite{Kramer1993}
\begin{equation}
    \Lambda =\left[\frac{W^2}{105\, \Omega^2}\right]^{-1}.
    \label{eq:intro-localization-length-zero-energy}
\end{equation}
Importantly, $\Lambda$ decreases with the on-site disorder $W$.
We note that the local disorder on the effective lattice is correlated in the lattice sites, which alters the expected localization length.
The reason being that spin domains share their atoms, and therefore the phonon states with all smaller domains.
Therefore, we calculate the connected correlation given by (see Supplementary Material for a derivation)
\begin{equation}
    C_m(l) = \braket{\hphon(m)\hphon(m+l)}_{c}.
    \label{eq:energy-correlation-function}
\end{equation}
For moderate distances $l\lesssim10$, the correlation approximately follows a power law with exponent $(-\alpha)$.
Physically, the longer the range of the correlation, the less of an ``effective'' disorder the particle experiences and therefore, the larger the localization length becomes.
Hence, there is a competition between the on-site disorder $W$ and the range of correlation (quantified by $\alpha$), which both increase with $\kappaone$ but have opposite effects on the localization length.
For a strength of the correlated disorder that decays as a power law the localization length is given by \cite{Croy2011}
\begin{equation}
    \Lambda_c(k) = \left[\frac{1}{8\sin^2(k)}\frac{W^2}{12}S(2k)\right]^{-1}
    \label{eq:localization-length-correlated}
\end{equation}
With $S(k)$ being the real part of the Fourier Transform of the correlation function, for which, in \cite{Makse1996}, an approximation 
was given, that reads
\begin{equation}
    S(k) = \frac{2\sqrt{\pi}}{\Gamma(\beta+1)}\left(\frac{k}{2}\right)^\beta\mathrm{K}_\beta(k)
    \label{eq:correlation-bessel}
\end{equation}
with $\beta = \frac{\alpha -1}{2}$, the gamma function $\Gamma$, and $\mathrm{K}_\beta$ being the modified Bessel function of order $\beta$.
Most notably, the localization length is here a function of the quasi particle's momentum $k$.
Since for large values of $m$, the correlation function $C_m(l)$ is symmetric in $l$, the spectral function is actually only a function of the modulus of $k$, i.e $S(k)=S(|k|)$, which translates to the localization length $\Lambda_c(|k|)$. 
$\Lambda_c(\vert k\vert)$ is depicted in Fig.~\ref{fig:loc-length} (b).
Since the wavefunction of a localized particle is a uniform superposition of all $k$-modes, the expectation value $\braket{|k|} = \pi/2$.
As can be seen in Fig.~\ref{fig:loc-length} (b) the corresponding analytical localization length has the highest agreement with the numerical results.
\\
With these results at hand, it is also possible to explain the transition from Bloch oscillations to Anderson Localization for increasing $\kappa$:
From scaling theory, it is generally believed that localization exists in 1D for arbitrarily small disorder strengths \cite{Kramer1993}.
In a finite system however, localization lengths larger than the system size (in our case $N$=100) cannot be observed.
From Fig.~\ref{fig:loc-length} (b), we see that the analytical (and numerical) localization length approaches this value at around $\kappaone /\Omega\approx3$. 
In our case, Bloch oscillations additionally limit dispersion of the particle to a smaller range than the localization length would permit, thereby shifting the point of observable localization to even larger values of $\kappaone$.
The transition to the localized state therefore takes place in our case when the localization length becomes of the order of the oscillation amplitude.

\FloatBarrier

\section{Phonon Dynamics for Large Domains}
%
\begin{figure}
    \centering
    \includegraphics[width=0.6\linewidth]{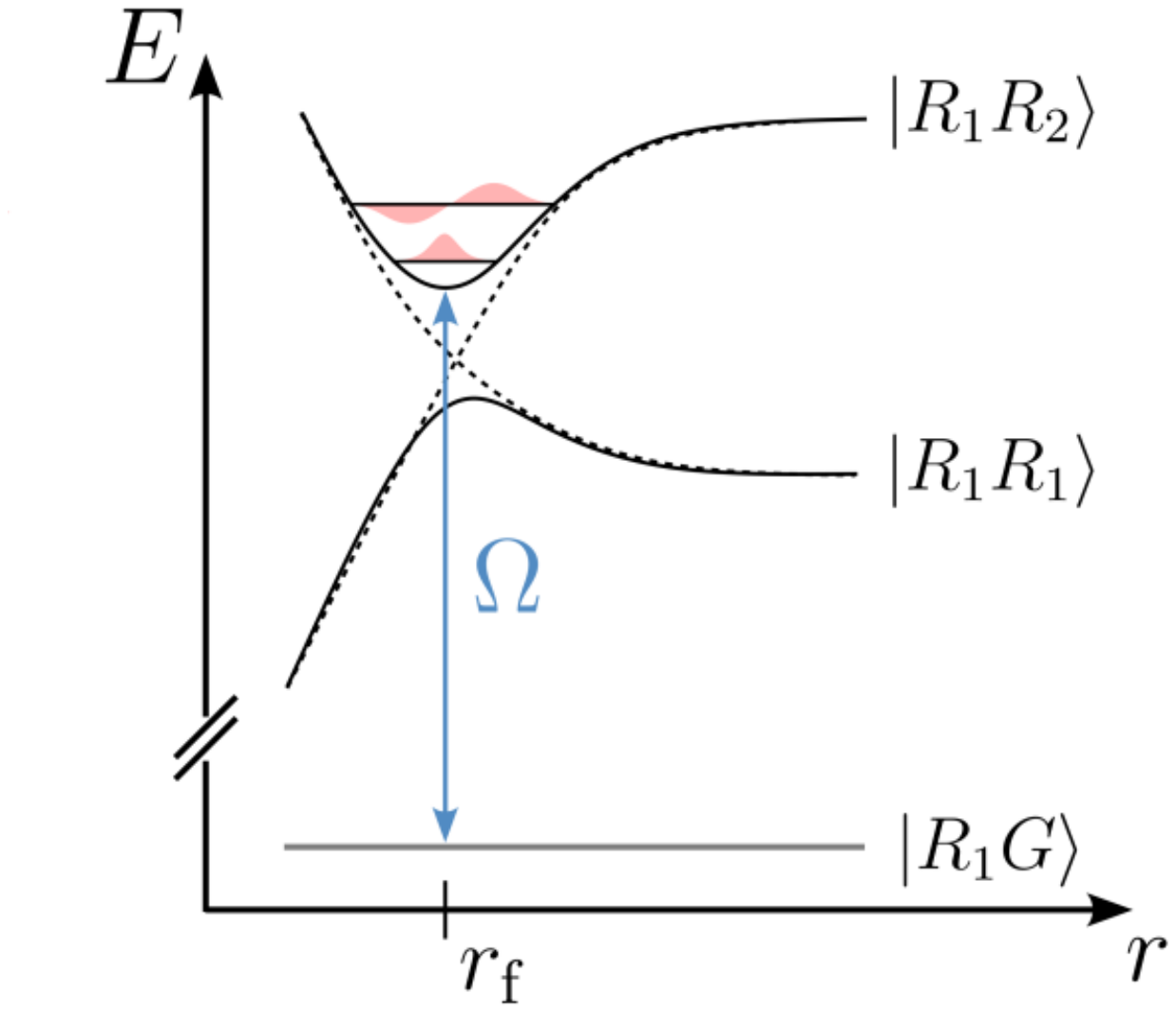}
    \caption{Two atom energy spectrum depending on distance $r$. Laser coupling with Rabi frequency $\Omega$ to an avoided crossing of attractive and repulsive Rydberg-Rydberg interaction potentials creates a nearly harmonic potential.}
    \label{fig:avoided-crossing}
\end{figure}

\begin{figure}[h]
  \centering
  \includegraphics[width=\columnwidth]{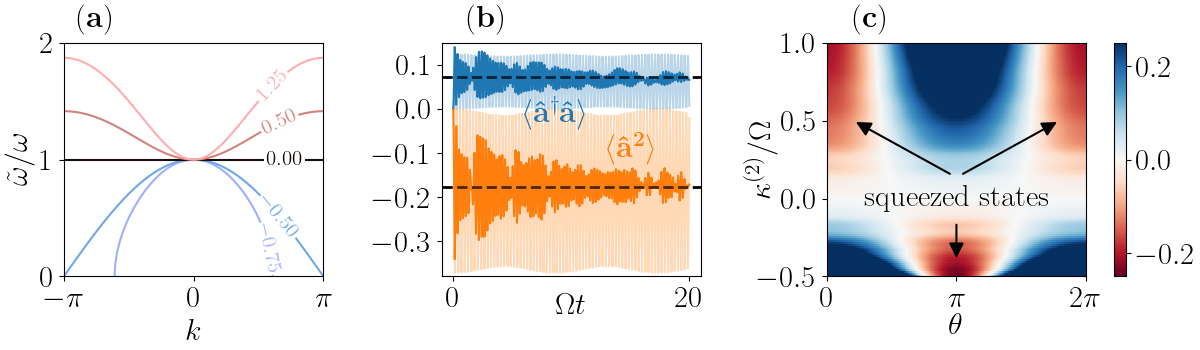}
  \caption{\textit{Phonon Dynamics:} (a) Dispersion relation from eq.~\eqref{eq:dispersion_relation} for different values of ${\kappatwo / \Omega}$ (inline numbers). Here, ${\kappatwo / \Omega = -0.5}$ corresponds to the critical coupling strength $\kappa_c$ and ${\kappatwo / \Omega = -0.75}$ is in the unstable phase as $\tilde \omega$ becomes imaginary for ${|k| \gtrsim 0.6 \pi}$. (b) Dynamics of expectation values of local oscillators $\opa_j$  in center of spin domain of size ${m=9}$ and for ${\kappatwo = \Omega}$. The faint blue and orange lines are the corresponding values without phonon transport, i.e. setting ${\hat{T}_{j,j+1} = 0}$ in Hamiltonian~\eqref{eq:lennard-jones-phonon-hamiltonian}. (c) Variance of generalized position operator minus vacuum variance, i.e. $\braket{\Delta Q_j^2} - 1$, with ${\hat Q_j = \mathrm{e}^{i\theta} \opadag_j + \mathrm{e}^{-i\theta} \opa_j}$ for site $j$ in center of cluster of size ${m=9}$, with ${\kappatwo = \Omega}$, and at time $\Omega t = 20$. For positive couplings, ${\kappatwo > 0}$, the system exhibits squeezed position states and for negative couplings, ${\kappatwo < 0}$, the system exhibits squeezed momentum states.
  }
\label{fig:phonon_dynamics}
\end{figure}

\textit{Facilitation to an avoided crossing of Rydberg potentials -- }
In this section, we want to investigate the phonon behavior ``deep'' within a spin domain.
While mechanical forces resulting from the van-der-Waals (vdW) potential, ${V(\hat{r}_{ij}) \sim \hat{r}_{ij}^{-6}}$ \cite{gallagher1994rydberg},
completely cancel out inside a defect free spin domain, they are highly relevant at the edges. This 
phonon source at the borders
can make the spin domain unstable and severely affect the facilitation dynamics \cite{magoni2024coherent}. 
This effect can be substantially reduced by considering an interaction potential resulting from
 an avoided crossing of two high-lying Rydberg states (see Fig.~\ref{fig:avoided-crossing})
(cf.~\cite{hollerith2019quantum} for an experimental realization).
In this case the interaction potential is approximately harmonic, and can even be inverted when laser coupling to the lower avoided crossing level. 
As opposed to the vdW potential, there is no force at the equilibrium position in the resulting Lennard-Jones potential. 
Assuming only nearest neighbor interactions, we can then write the Rydberg-Rydberg interaction to second-order as ${V(\hat{r}_{j, j+1}) \approx V(\rfac) + \frac{1}{2}V^{\prime \prime}(\rfac) (\hat{r}_{j, j+1} - \rfac)^2}$. 
Hence, the resulting problem is the same as discussed in the previous section, however with the absence of the linear, phonon creating term.
The phonon Hamiltonian for this problem is given by
\begin{equation}
\begin{aligned}
\hphon(m) 
&= \sum_{l=0}^{m-1}\omega (\opadag_l \opa_l + \tfrac{1}{2}) + \kappatwo \sum_{l=0}^{m-2}\bigl(\hat S_l + \hat S_{l+1}- 2 \hat T_{l, l+1}\bigr)
\end{aligned}
\label{eq:lennard-jones-phonon-hamiltonian}
\end{equation}
with  spin-phonon coupling 
${\ktwo = V^{\prime \prime} (\rfac) / 4 m \omega}$, which, in the absence of a first order derivative, we can treat as an independent parameter. 
$\kappatwo$ is positive (negative) when coupling to the upper (lower) avoided crossing level, see Fig.~\ref{fig:avoided-crossing}.
Note that the quadratic and transport terms, not present in the linear approximation of strictly local phonons, assumed in \cite{magoni2024coherent}, are crucial as they lead to non-classical correlations in the atomic positions and to a fast thermalization of the motional degrees of freedom in the dynamics from a general initial state.
In this section, we use for numerical simulation $\omega = 8\Omega$ and $\vnn = 500\Omega$.
\\
As the phonon Hamiltonian Eq.~\eqref{eq:lennard-jones-phonon-hamiltonian} is formally a special case of Eq.~\eqref{eq:full-phonon-hamiltonian} for $\kappaone= 0, \kappatwo\neq0$, we can diagonalize it with the same method used in the previous section.
The phonon operators for which $\hphon$ becomes diagonal are the $\opd_q$ operators from Eq.~\eqref{eq:definition_opd}.
Assuming a sufficiently large spin domain, we can introduce continuous fields $\hat {d}(k) = \lim_{m\to\infty}\hat {d}_k \sqrt{\frac{m}{\pi}}$ and find 
\begin{align}
    \label{eq:capital_d_integral}
     \hat h_p(m)
    \approx
    \int_0^{\pi}\!\!\! \mathrm{d}k \; \tilde \omega(k) 
       \Bigl( \hat{d}^\dagger (k) \hat{d}(k) + \frac{m}{2\pi} \Bigr).  
\end{align}
The phonon dispersion is given by 
\begin{align}
    \label{eq:dispersion_relation}
    \tilde \omega(k) = 
    \sqrt{
        \omega^2 + 8 \omega \ktwo 
        \big(1 - \cos k\big)
    }.
\end{align}
and plotted in Fig.~\ref{fig:phonon_dynamics}a (The dispersion relation like any other quantity gained by the Discrete Cosine Transformation is defined on the interval $k\in [0, \pi)$ and assumed to be symmetric around $k=0$. For a more intuitive interpretation of the results, we plot the dispersion relation in the interval $k\in [-\pi, \pi]$.)
From the dispersion relation we can readily see that the phonons become unstable at the critical coupling strength ${\kappatwo_c \equiv -\frac{\omega}{16}}$. 
(Since we use ${\omega = 8 \Omega}$ for all simulations, this corresponds to ${\kappatwo_c = -\frac{1}{2} \Omega}$.)
At ${\kappatwo_c}$ we find a mode softening for ${k=\pm \pi}$.
For $\kappatwo > \kappatwo_c$ the ground state of the phonon Hamiltonian is the vacuum state in all $\opd_k$, corresponding to a correlated squeezed vacuum of the local oscillators $\opa_{j}$. 

%

\ 

\textit{Phonon-thermalization --} Since in any realistic experiment, the ground state of the total Hamiltonian cannot easily be prepared, we  consider a system initially in a finite spin domain of length $m$ and all atoms are in the ground state of the local oscillators $\opa_j$. As shown in Fig.~\ref{fig:phonon_dynamics}b due to the phonon transport, local observables quickly approach a stationary value in the center of a large spin chain.

The long-time expectation values of the local phonon operators within the spin domain, i.e. $\braket{\opadag \opa}$ and $\braket{\opa^2}$,
can be calculated as follows:
As the phonon Hamiltonian is diagonal in the $\opd$ basis, $\braket{\opddag \opd}$ is constant, 
and $\braket{\opd^2}_t = \braket{\opd^2}_0 \mathrm{e}^{-2i\tilde \omega t}$. Therefore, the time averaged values of $\braket{\opadag \opa}$ and $\braket{\opa^2}$ can be obtained by expressing these operators in the $\opd$ basis and neglecting $\braket{\opd^2}$ terms. We obtain the time averaged operator values (for details see Supplementary Material) as
\begin{subequations}
    \label{eq:phonon_analytics}
    \begin{align}
        \label{eq:n_a_analytic}
        \braket{\opadag \opa} &= +\frac{\kappatwo}{\omega} + \frac{1}{8}
        \Bigg(
            \sqrt{\frac{\omega}{\omega + 16\kappatwo}} - 1
        \Bigg)
        \\
        \label{eq:a_sq_analytic}
        \braket{\opa^2} &= -\frac{\kappatwo}{\omega} + \frac{1}{8}
        \Bigg(
            \sqrt{\frac{\omega}{\omega + 16\kappatwo}} - 1
        \Bigg).
    \end{align}
\end{subequations}

In Fig.~\ref{fig:phonon_dynamics}b we have plotted these values as dashed lines along with the time evolution of local phonon correlations inside the domain including (dark lines) and excluding (faint lines) transport, obtained by TEBD simulations.
We see a very good agreement when transport is taken into account in the numerics. Without the transport terms, i.e. ${\hat{T}_{j, j+1} =0}$, the local phonon operators oscillate, 
whereas, in the presence of phonon transport these quantities thermalize despite the pure unitary evolution of $\hat H$, following the eigenstate thermalization hypothesis (ETH) \cite{deutsch1991quantum, srednicki1994chaos}.

\ 

\textit{Phonon squeezing -- } 
Finally, we want to quantify the quantum fluctuations present in local oscillators, as a result of dipolar interactions. To this extent, we determine the variance of a generalized position operator $\braket{\Delta \hat Q^2}$, defined as \cite{mandel1995optical} $\hat Q = \mathrm{e}^{i\theta} \opadag + \mathrm{e}^{-i\theta}\opa$,
where the angle $\theta$ allows us to sample $(q, p)$ phase space. Under this convention, the vacuum value is ${\braket{\Delta \hat Q^2}_\mathrm{vac} = 1}$. In Fig.~\ref{fig:phonon_dynamics}c, ${\braket{\Delta \hat Q^2} - 1}$ is plotted over the coupling $\kappatwo$ and the angle $\theta$. In the figure, negative values correspond to states which are more strongly localized than vacuum fluctuations, i.e. squeezed states.

\begin{figure}
  \centering

  \includegraphics[width=\columnwidth]{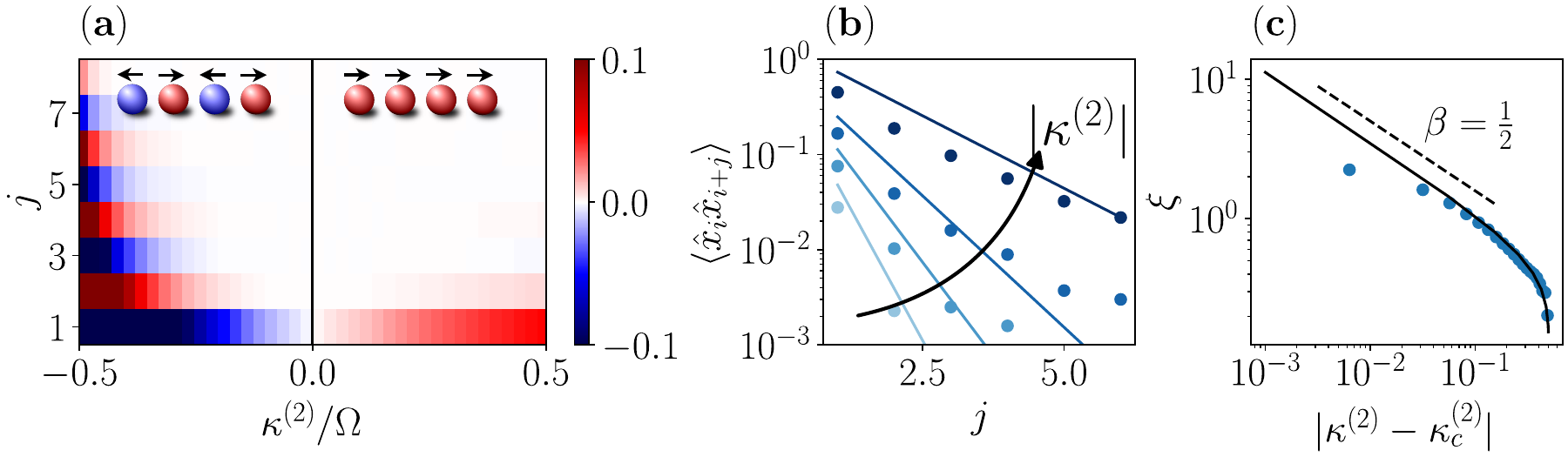}
  \caption{\textit{Phonon Correlations:} (a) Time-averaged displacement correlation ${C_{ij} \equiv \braket{\hat{x}_i \hat{x}_{i+j}}}$ from TEBD simulations for ${i=-3}$ and an initial spin domain size ${m_0 = 9}$ (${i=0}$ is the domain center). A positive product (red) signifies phonons oscillating in phase and a negative product (blue) corresponds to out of phase oscillations. (b) Spatial correlation of $i,j$ for $\kappatwo/\Omega=[-0.44,-0.34,-0.24,-0.14]$ and for $i = -3$ from numerics (dots), as well as analytically calculated correlation $C_{ij}$ from eq.~\eqref{eq:c_ij} (solid lines). (c) Correlation length $\xi$ taken from exponential fit of decay of ${C_{ij}}$ with distance between $(i, j)$ from numerics (dots) and analytics (solid black line). The correlation length extracted from numerics saturates due to the finite size of the spin domain. Approaching $\kappa_c$, $\xi$ diverges yielding the mean field exponent ${\beta = \frac{1}{2}}$.}
\label{fig:phonon_correlations}
\end{figure}

For positive couplings, i.e. $\kappatwo > 0$, we find squeezed position states, i.e. the variance is minimal for ${\theta = 0}$. 
This results from atoms being subject to the tweezer trapping potential and an additional trapping potential emerging from the Rydberg-Rydberg interaction.
For negative couplings, i.e. ${\kappatwo < 0}$, the Rydberg-Rydberg interaction potential is inverted and the system converges toward a mode softening at criticality at ${k = \pm \pi}$. 
Consequently, momentum states become squeezed.

\ 

\textit{Correlated Atomic Motion -- } Having discussed the local behavior of phonons we now turn to their correlations. 
To this extent, we can regard the displacement correlation $C_{ij} \equiv \braket{\hat x_i \hat x_{i+j}}$. For positive values of $C_{ij}$, phonons $i$ and $i+j$ displace from equilibrium in the same direction, whereas for negative values of $C_{ij}$ they displace in opposite directions. Within spin domains we find strong phonon-phonon correlations, which become long-range near $\kappatwo_c$.
Moreover, we find the oscillatory behavior of phonons in the domain to change qualitatively from in-phase correlations
(${\kappa^{(2)} > 0}$) to out-of-phase correlations (${\kappatwo < 0}$) (see Fig.~\ref{fig:phonon_correlations}a).

Using ${\hat x_i = \opadag_i + \opa_i}$, we can express $C_{ij}$ in terms of $\opd$-basis phonons. Neglecting the oscillating terms $\opd^2$, this yields 

\begin{align}
    \label{eq:c_ij}
    C_{ij} &= \frac{1}{\pi} \int_0^{\pi} \mathrm{d}k \;\cos(kj)\bigl[(u+v)^2\,(u^2+v^2)\bigr]    
\end{align}

where $u$ and $v$ are $k$ dependent (see text above). 
The solution to this integral yields an exponential decay of $C_{ij}$ with the distance  $j$ for ${|\kappatwo - \kappatwo_c| > 10^{-4}}$ (solid lines in Fig.~\ref{fig:phonon_correlations}b). From this we can extract a correlation length $\xi$ using ${C_{ij} \sim \mathrm{e}^{-\frac{j}{\xi}}}$. This correlation length diverges near the critical point $\kappatwo_c$ and scales as 
\begin{align}
    \xi \sim {|\kappatwo - \kappatwo_c|^{-\beta}},
\end{align}
with critical exponent $\beta$ consistent with the mean-field value ${\beta = \frac{1}{2}}$ (cf.~Fig.~\ref{fig:phonon_correlations}c).

\section{Summary}

In summary, we discussed the coupled excitation and motional dynamics of a chain of  Rydberg atoms trapped in tweezer arrays and laser-driven under the facilitation constraint.
Here excitation spreading is caused by the dipole-dipole interactions between atoms in the Rydberg state, which however also couple internal degrees of freedom (spin) to atomic motion (phonons) by mechanical forces on the atomic positions.
When spontaneous decay can be neglected, the facilitation dynamics reduces to the dynamics of domains of contiguous Rydberg excitations which grow or shrink at the edges but neither coalesce nor split.
We studied the effect of spin-phonon coupling on the dynamics of such domains and analyzed the buildup of correlations in the motion of the atoms.
In the bulk of an excitation domain dipole forces from  adjacent Rydberg atoms compensate and lead to a harmonic confinement potential.
Expanding the van-der-Waals interaction potential to second order around the lattice spacing yields a smallness parameter $\kappa=r_\textrm{osc}/r_\textrm{fac}$, given by the ratio of the oscillator length and the facilitation distance, that serves as a measure for the spin-phonon coupling strength. 
Numerical simulations showed that tuning this single parameter  
at moderate coupling strengths, Bloch oscillations emerge, which prevent the spin domain from expanding ballistically.
Increasing the coupling strength further results in an Anderson-like localization of the domain size. To explain these effects of spin-phonon coupling,
we first showed that the many-body excitation dynamics can be mapped to a single particle hopping on a semi-infinite lattice, where the lattice position corresponds to the domain size. Spin-phonon coupling
then leads to phonon-induced local potential energies. 
The mean-field contribution of these energies corresponds to a linear potential gradient in the effective domain model, causing Bloch oscillations. For stronger spin-phonon couplings, $\kappa$, quantum fluctuations in the phonons become relevant and result in correlated potential disorder in the effective domain model, which explains the observed localization. 

Subsequently, we investigated the motional state of the atoms within a domain considering a tailored  interaction potential arising from 
laser excitation to an avoided level crossing of two Rydberg states.
In the resulting Lennard-Jones-type potential, 
in which there are no average mechanical forces, spin-phonon coupling 
leads to an effective Hamiltonian, that contains only quadratic terms in the phonon operators. These terms describe pair creation and annihilation as well as phonon transport. The latter leads to fast thermalization of phonons and the emergence of strong correlations of motional states of the atoms. For a repulsive Lennard-Jones potential
the displacements of the atoms from their equilibrium positions show an oscillating (``anti-ferromagnetic'') correlation with finite correlation length. The correlation length diverges when the spin-phonon coupling approaches a critical value, where the system becomes mechanically unstable. Close to this critical point the pair creation and annihilation terms in the phonon Hamiltonian lead to strongly squeezed phonon states.

%
%

\ack{
The authors thank M. Magoni, D. Petrosyan and J. Sirker for fruitful discussions.}

\funding{
Financial support from the DFG through SFB TR 185, Project No. 277625399 is gratefully acknowledged. The authors also thank the Allianz für Hochleistungsrechnen (AHRP) for giving us access to the ``Elwetritsch'' HPC Cluster.}

\roles{
MF conceived, conceptualized and supervised the project and contributed to the theoretical understanding of the numerical results. BB developed the theoretical understanding of the localization and carried out the corresponding numerical studies. DB contributed to the conceptualization of the project, wrote the original simulation software, and together with BB, carried out the investigations. 
He also developed the analysis of the phonon dynamics.
All three authors contributed to writing, reviewing, and editing the manuscript.
}

\data{Numerical simulation data will be made available at a later point.}

\suppdata

\subsection{Exponential localization of wavefunction}
\begin{figure}[h]
    \centering
    \includegraphics[width=0.9\linewidth]{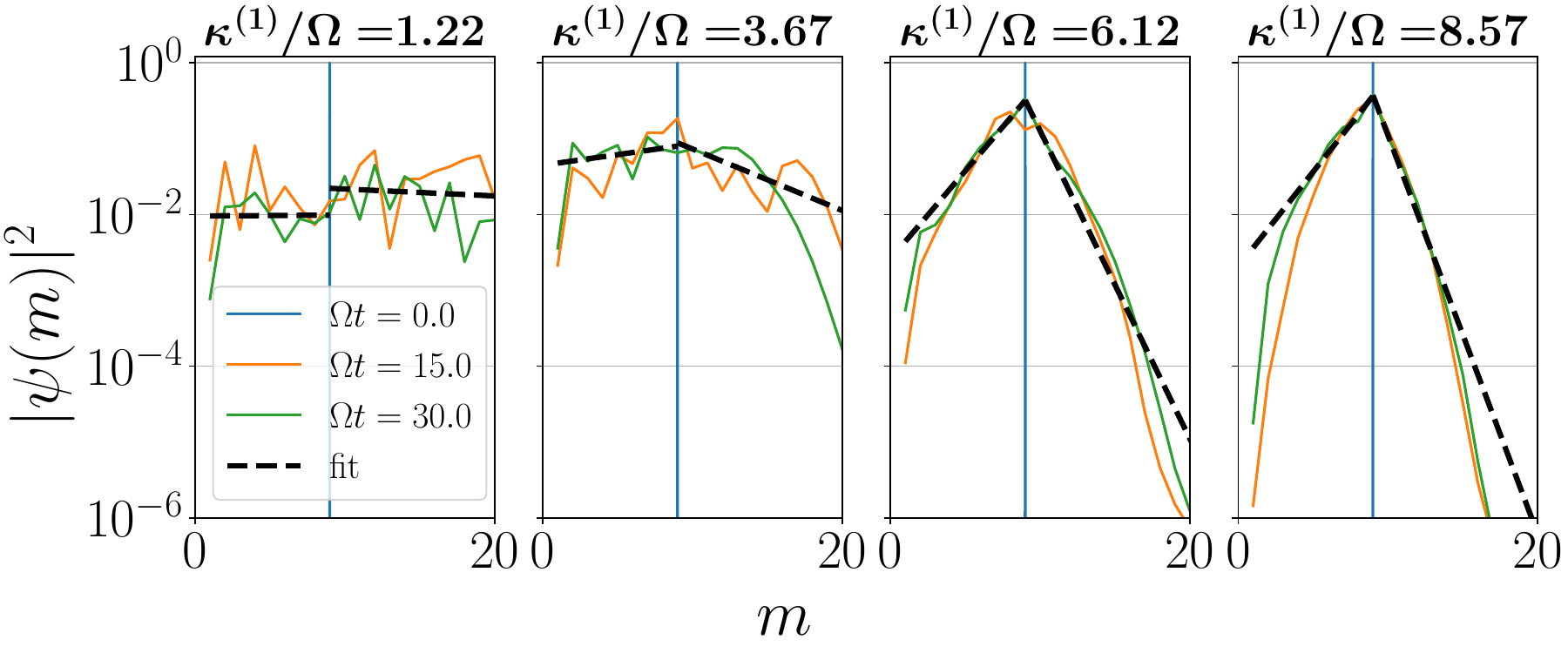}
    \caption{Single particle wavefunction $\psi(m)$ in spin-domain space at different time steps. The wavefunction becomes exponentially localized around the initial state $m_0 = 9$ for large values of $\kappaone$. Localization lengths to the left and right of the center vary slightly due to the ``hard wall'' at $m=1$. For the numerical study, we therefore use only the right flank of the wavefunction.}
    \label{fig:exponential-localization}
\end{figure}

The simulation data from the unidirectionally expanding spin domain can be used to compare numerical results with the analytic prediction of localization in the effective Hamiltonian by projecting the numerical wavefunction to spin domain space via eq.~\eqref{eq:m-projector}. 
We plot the single particle wavefunction $\psi(m)$ for different time steps in Fig.~\ref{fig:exponential-localization}, which clearly shows that the wavefunction becomes exponentially localized.

\subsection{Diagonalization of the Phonon Hamiltonian}

In the following we diagonalize the phonon Hamiltonian under the Born-Oppenheimer approximation for a fixed domain length $m$.  It is given by
\begin{eqnarray}
\hphon(m) 
&=& \sum_{l=0}^{m-1}\omega (\opadag_l \opa_l + \tfrac{1}{2})
  + \sum_{l=0}^{m-2}\bigl( \hat{{V}}_{l,l+1} - \vnn\bigr) \label{eq:full-phonon-hamiltonian}\\
&=& \sum_{l=0}^{m-1}\omega (\opadag_l \opa_l + \tfrac{1}{2})  - \kappaone \sum_{l=0}^{m-2}(\opadag_{l+1}+\opa_{l+1}-\opadag_l-\opa_l)  + \kappatwo \sum_{l=0}^{m-2}(\opadag_{l+1}+\opa_{l+1}-\opadag_l-\opa_l)^2,
\nonumber
\end{eqnarray}
where we introduce the shorthand notation 
$\kappaone =V'(\rfac)\rosc =  6 \vnn \frac{\rosc}{\rfac}$ 
and
$\kappatwo = \frac{1}{2} V''(\rfac)\rosc ^2 = 21 \, \vnn \frac{\rosc ^2}{\rfac^2} $.
The second term in the last line represents a telescoping series, meaning all terms cancel except those at sites $l=0,m-1$.
Writing out the sum in the last term, we can recast the Hamiltonian in the following matrix representation:
\begin{equation}
    \begin{split}
          \hphon(m) = 
          & \begin{pmatrix}
        \opadag_0 \\
        \opadag_1 \\
        \vdots \\
        \opadag_{m-2} \\
        \opadag_{m-1}
    \end{pmatrix}^T
    \cdot
    \left[
        \omega \mathbbm{1}_m+ 2\kappatwo
    \underbrace{\begin{pmatrix}
            1 & -1 & &  & \\
            -1 & 2 & -1 & &  \\
            &\diagdown&\diagdown&\diagdown& \\
             & & -1 & 2 & -1\\
             &&&-1&1\\
        \end{pmatrix}}_{A}   \right]\cdot
    \begin{pmatrix}
        \opa_0 \\
        \opa_1 \\
        \vdots \\
        \opa_{m-2} \\
        \opa_{m-1}
    \end{pmatrix}\\
    & +\kappatwo \begin{pmatrix}
        \opadag_0 \\
        \opadag_1 \\
        \vdots \\
        \opadag_{m-2} \\
        \opadag_{m-1}
    \end{pmatrix}^T
    \underbrace{\begin{pmatrix}
            1 & -1 & &  & \\
            -1 & 2 & -1 & &  \\
            &\diagdown&\diagdown&\diagdown& \\
             & & -1 & 2 & -1\\
             &&&-1&1\\
        \end{pmatrix}}_{A}
         \begin{pmatrix}
        \opadag_0 \\
        \opadag_1 \\
        \vdots \\
        \opadag_{m-2} \\
        \opadag_{m-1}
    \end{pmatrix} + \hc \\
    & - \kappaone
  (\opadag_{m-1} + \opa_{m-1}
  - \opadag_{0} - \opa_{0}) + \mathrm{const.}
    \end{split}
\label{eq:phonon-hamiltonian-matrix}
\end{equation}
The orthonormal transformation that diagonalizes matrices of the type of $A$ is the Discrete Cosine Transform (DCT) type $\mathrm{II}$ \cite{britanak2007dct}:
\begin{align}
\label{eq:dct-transformed-ops}
    \opa_l &= \sqrt{\frac{2}{m}}\sum_{q=0}^{m-1}\epsilon_q \Cos{\frac{\pi q}{m}(l + \frac{1}{2})} \opcapitala_q,\qquad  \opcapitala_q = \sqrt{\frac{2}{m}}\sum_{l=0}^{m-1}\epsilon_l \Cos{\frac{\pi l}{m}(q + \frac{1}{2})} \opa_l
\end{align}
with $\epsilon_0 = 1/\sqrt{2}$ and $\epsilon_{q\ne 0} =1$.
Inserting eq.~\eqref{eq:dct-transformed-ops} into eq~\eqref{eq:phonon-hamiltonian-matrix}, we get
\begin{equation}
    \begin{split}
        \hphon(m) = & \sum_{q=0}^{m-1}\underbrace{(\omega + 2 \gamma_q)(\opcapitaladag_q\opcapitala_q + \frac{1}{2}) + \gamma_q(\opcapitaladagsq_q + \opcapitalasq_q)}_{\hat{h}_q^{(2)}} \\
        + & \sum_{q=0}^{m-1} \underbrace{\kappaone \sqrt{\frac{2}{m}}\epsilon_q \opcapitaladag_q \left[ \Cos{\frac{\pi q}{2m}} - \Cos{\frac{\pi q}{m}(m - \frac{1}{2})}\right] + \hc}_{\hat{h}_q^{(1)}}  + \mathrm{const}  \;.
    \end{split}
    \label{eq:full-hamiltonian-dct}
\end{equation}
with
\begin{equation}
    \gamma_q = 2\kappatwo(1 - \costerm).
    \label{eq:definition_gammaq}
\end{equation}
To eliminate the squeezing terms we introduce the Bogoliubov transformed bosonic operators
\begin{equation}
    \opd_q = u_q\opcapitala_q -  v_q\opcapitaladag_q
    \label{eq:definition_opd}
\end{equation}
with $u_q = \cosh{\theta_q}$ and $v_q = \sinh{\theta_q}$ and the angle 
\begin{equation}
    \theta_q = \frac{1}{2}\arctan\left(-\frac{2\gamma_q}{\omega + 2\gamma_q}\right).
    \label{eq:definition_theta}
\end{equation}
After inserting Eq.~\eqref{eq:dct-transformed-ops} in the above expression, one obtains the following Hamiltonian ${\hat h_q = \hat h_q^{(1)} + \hat h_q^{(2)}}$, representing a displaced harmonic oscillator:
\begin{equation}
    \frac{h_q}{\widetilde{\omega}_q} = \opddag_q\opddag_q + \frac{1}{2} + \underbrace{\frac{\alpha_q(u_q+v_q)}{\omegatildeq}}_{\mu_q}(\opddag_q + \opd_q).
    \label{eq:definition_muq}
\end{equation}
with 
\begin{equation}
    \alpha_q = \kappaone \sqrt{\frac{2}{m}}\epsilon_q \left[ \Cos{\frac{\pi q}{2m}} - \Cos{\frac{\pi q}{m}(m - \frac{1}{2})}\right].
\end{equation}
The linear displacement terms are eliminated by the canonical transformation
\begin{equation}
        \opg_q = \opd_q + \mu_q.
        \label{eq:definition_opg}
\end{equation}
The diagonalized phonon Hamiltonian then finally reads
\begin{equation}
    \hphon(m) = \sum_{q=0}^{m-1}\omegatildeq\Bigl(\opgdag_q\opg_q + \frac{1}{2}\Bigr) + \delta(m).
\label{eq:full-g-ham}
\end{equation}
with 
\begin{equation}
\delta(m) = -\sum_{q=0}^{m-1} \mu_q^2\omegatildeq    
\label{eq:definition_delta}
\end{equation}

\subsection{Expectation value in initial state}

For sufficiently large domain sizes $m$, the ground state energy shift $\delta(m)\propto \frac{1}{m}$ becomes negligible.
At the same time, the increments between adjacent values of $k= \frac{\pi q}{m}$ become small enough to justify a continuum approximation:
\begin{equation}
\begin{split}
   & \frac{1}{2}\sum_{q=0}^{m-1}\omegatilde_q  = \frac{1}{2}\sum_{q=0}^{m-1}\sqrt{\omega^2 + 8\omega\kappatwo\left(1 - \cos\left(\frac{\pi q}{m}\right)\right)}\\
   &\qquad \, \quad \overset{m\to \infty}{=}\frac{m}{2\pi}\int_0^{\pi}\mathrm{d}k \underbrace{\sqrt{\omega^2 + 8\omega\kappatwo(1 - \cos(k))}}_{\omegatilde(k)}
\end{split}
\end{equation}
We replace the operators $\opg_k$ by normalized field operators
\begin{equation}
    \opg_k \to \sqrt{\frac{\pi}{m}}\opg(k)
    \label{eq:opg-continuum-approximation}
\end{equation}
such that $\sum_k\opgdag_k\opg_k = \int\mathrm{d}k\opgdag(k)\opg(k)$.
The expectation value 
\begin{equation}
    \langle\hphon(m)\rangle = \int_0^{\pi}\mathrm{d}k\;\omegatilde(k)\langle\opgdag(k)\opg(k)\rangle +\frac{m}{2\pi}\int_0^{\pi}\mathrm{d}k \; {\omegatilde(k)}
\end{equation}
is now divided into two terms.
We note that $\opgdag(k)\opg(k)$ is a constant of motion of the Hamiltonian and independent of $m$.
Therefore, only the second summand in the above equation provides a contribution to the lattice potential that is not constant in $m$.
We can further simplify the second summand using the complete elliptic integral $\elliptic$ \cite{wolfram_2026_elliptice}:
\begin{equation}
\frac{m}{2\pi}\int_0^{\pi}\mathrm{d}k \; {\omegatilde(k)}
    = \frac{m}{2}\underbrace{\frac{1}{\pi}\sqrt{\omega^2 + 16\omega\kappatwo}\elliptic\left(\frac{16\omega\kappatwo}{\omega^2 + 16\omega\kappatwo}\right)}_{\langle{\omegatilde}\rangle}.
\end{equation}

\subsection{Uncertainty and Disorder Correlation}

\begin{figure}
    \centering
    \includegraphics[width=\linewidth]{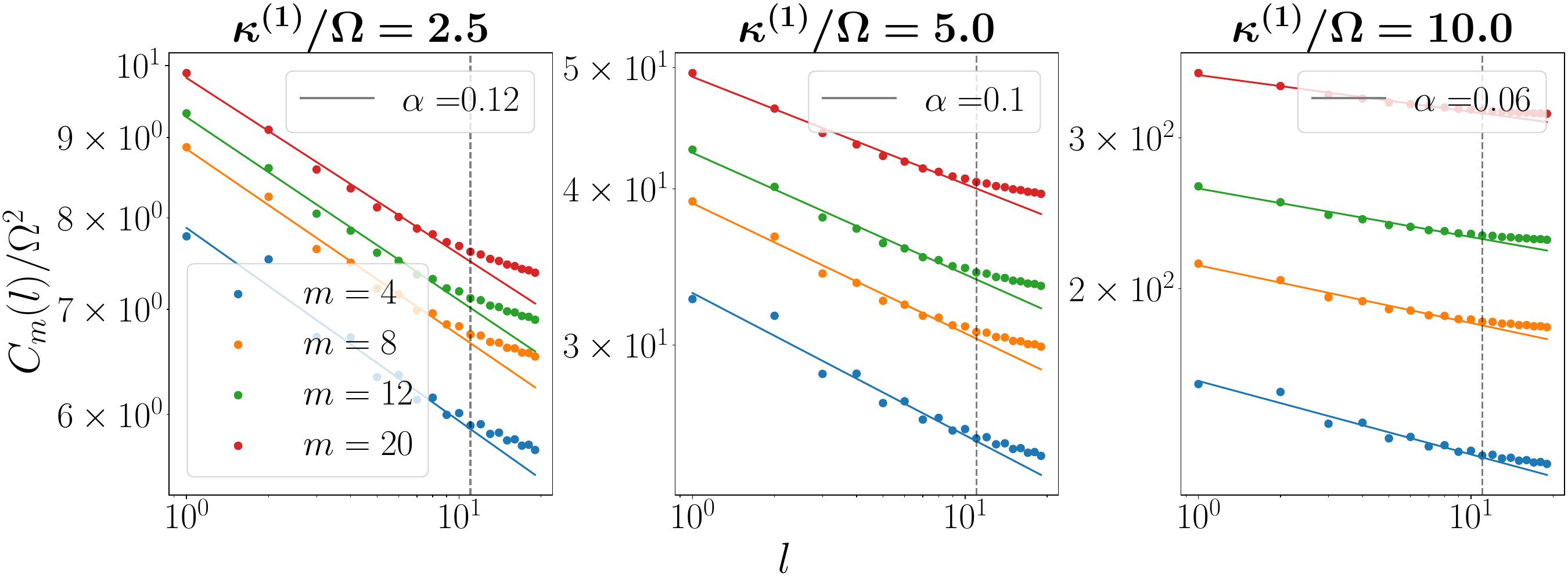}    
    \caption{Correlation of phononic energies Eq.~\eqref{eq:energy-correlation-function} of different domain sizes as a function of ``distance'' $l$.}
    \label{fig:hphon-correlation}
\end{figure}

Here, we evaluate the expression for the variance of the phonon energy in the initial state, which is the vacuum of the $\opa$ phonons, i.e.
\begin{equation}
    \Delta\hphon(m)^2 = \braket{\hphon(m)^2} - \braket{\hphon(m)}^2.
\end{equation}
Terms in $\hphon(m)$ which are $c$-numbers cancel in the variance, so the above expression can be reduced to
\begin{equation}
    \Delta\hphon (m)^2= \sum_{q, q'}^m\omegatildeq\omegatilde_{q'}\Bigl(\braket{\opgdag_q\opg_q\opgdag_{q'}\opg_{q'}} - \braket{\opgdag_q\opg_q}\braket{\opgdag_{q'}\opg_{q'}}\Bigr)
\end{equation}
The second summand in this equation, the quadratic correlations, can be calculated straightforwardly by inserting Eqs. \eqref{eq:definition_opg} and \eqref{eq:dct-transformed-ops}, which yields 
\begin{equation}
    \begin{split}
        \braket{\opgdag_q\opg_q} &= \braket{(\opddag_q + \mu_q)(\opd_q + \mu_q)}\\
        &=\braket{(u_q \opcapitaladag_q - v_q \opcapitala_q + \mu_q)(u_q \opcapitala_q - v_q \opcapitaladag_q + \mu_q)}=v_q^2 + \mu_q^2.
    \end{split}
    \label{eq:mean-value-opg}
\end{equation}
The same can be done for the quartic correlations.
By additionally substituting $u_q$, $v_q$ and $\mu_q$ by their respective definitions, the variance can finally be reduced to
\begin{equation}
    \begin{split}
    (\Delta\hphon(m))^2= &\sum_{q, q'}^m\omegatildeq\omegatilde_{q'}\delta_{q, q'}\Bigl(\mu_q^2(u_q - v_q)^2 + 2(u_q v_q)^2\Bigr)= \sum_q^m\alpha_q^2 + 2\gamma_q^2.
    \end{split}
    \label{eq:variance_hphon}
\end{equation}
Finally, we want to calculate the correlation between the phononic energies of two different domain sizes $m$ and $m'$.
The calculation is analogous to that of the variance, except that we have to substitute 
\begin{equation}
    \delta_{q, q'}\to \braket{\hat{A}^{(m')}_{q'}\hat{A}^{\dagger (m)}_{q}}
\end{equation} 
in Eq.~\eqref{eq:variance_hphon}, where we introduced the superscript $(m)$ to distinguish between DCTs of different domain sizes. 
Physically, this quantity represents the overlap between two phonons states in different $q$-modes of different domain sizes. 
It is given by
\begin{equation}
\begin{split}
 \braket{\hat{A}^{(m')}_{q'}\hat{A}^{\dagger (m)}_{q}}
 &=\sqrt{\frac{2}{m}}\sqrt{\frac{2}{m'}}\sum_{j=0}^{m'-1}\sum_{l=0}^{m-1}\epsilon_j\epsilon_l\Cos{\frac{\pi j}{m'}\left(q'+\frac{1}{2}\right)}\Cos{\frac{\pi l}{m}\left(q+\frac{1}{2}\right)}   \underbrace{\braket{\opa_j\opadag_l}}_{\delta_{jl}}\\
 &=\sqrt{\frac{2}{m}}\sqrt{\frac{2}{m'}}\sum_{j=0}^{\min(m, m')-1}\epsilon_j^2\Cos{\frac{\pi j}{m'}\left(q'+\frac{1}{2}\right)}\cos{\frac{\pi j}{m}\left(q+\frac{1}{2}\right)}.
\end{split}
\label{eq:Phonon-A-correlation}
\end{equation}
For the same domain size $m=m'$, this expression reduces again to a Kronecker delta due to the Orthogonality of the DCT.
The correlation finally is given by
\begin{equation}
\begin{split}
    C_m(l) &= \braket{\hphon(m')\hphon(m)} - \braket{\hphon(m)}\braket{\hphon(m')} \\
    &= \sum_{q=0}^{m-1}\sum_{q'=0}^{m'-1}\alpha^{(m)}_q\alpha^{(m')}_{q'}\braket{\hat{A}^{(m')}_{q'}\hat{A}^{\dagger (m)}_{q}} + 2 \gamma^{(m')}_{q'}\gamma^{(m)}_{q}\braket{\hat{A}^{(m')}_{q'}\hat{A}^{(m)}_{q}}^2
\end{split} 
\label{eq:connected-correlation}
\end{equation}
which is plotted in Fig.~\ref{fig:hphon-correlation}. As can be seen, the correlation decreases approximately like a power law for moderate distances $l\lesssim10$. The exponent $\alpha$ is the same for all domain sizes $m$ and depends only on $\kappaone$.
Thus the potential caused by the vacuum fluctuations of the phonons corresponds to correlated disorder.

\subsection{Calculation of Phonon Observables}

We want to calculate the phonon observables in the original basis, $\braket{\opadag\opa}$, $\braket{\opa^2}$, and $C_{ij} \equiv \braket{\hat{x}_i \hat{x}_{i+j}}$. All of these can be obtained by expressing them in the $\opd$-basis and neglecting fluctuations, i.e. ${\braket{\opd^2} \approx 0}$. Since ${n_d \equiv \braket{\opddag \opd}}$ is constant, we can write ${n_d(t) = n_d(0)}$. For the initial condition where $\opa$-phonons are in the Fock~$\ket{0}$ state, all contributions $\braket{\opadag_j \opa_j}$ and $\braket{\opa^2_j}$ vanish at ${t=0}$, and the population of $\opd$ phonons is given by
\begin{align}
    n_d(t) = n_d(0)
    = v^2.
\end{align}
Expressing the above mentioned observables in $\opd$-phonons, with ${a = \frac{\pi}{m}}$, and neglecting fluctuations, we find in the homogeneous limit (i.e. $\braket{\opadag_j\opa_j}\to\frac{1}{m}\sum_i\braket{\opadag_j\opa_j}$)
\begin{subequations}
    \begin{align}
        \braket{\opadag_j \opa_j} &= \frac{1}{m}\sum_q (u_q^2 + v_q^2)v_q^2 + v_q^2 \equiv \frac{1}{m} \sum_q 2 u_q^2 v_q^2
        \\
        \braket{\opa_j^2} &= \frac{1}{m} \sum_q (u_q^2 + v_q^2)u_qv_q
        \\
        C_{ij} &= \frac{1}{m}\sum_q  \cos(qaj)\bigl[(u_q+v_q)^2\,(u_q^2+v_q^2)\bigr] ,
    \end{align}
\end{subequations}
where we used $u^2 - v^2 = 0$ and $j \ll m$ in the last line.
Assuming a sufficiently large cluster, ${m \gg 1}$, we can evaluate the sum by an integral approximation. 
Inserting ${u = \cosh \theta_c}$ and ${v = \sinh \theta_c}$, with $\theta_c$ given by eq.~\eqref{eq:definition_theta}, using trigonometric relations we receive the integrals

\begin{subequations}
    \begin{align}
        \braket{\opadag_j \opa_j} &= \frac{2}{\pi} \int_0^{\pi} \mathrm{d}k \;\frac{4(\kappatwo)^2(1 - \cos k)^2}{\omega^2 + 8 \omega \kappatwo(1 - \cos k)}
        \\
        \braket{\opa_j^2} &= \frac{1}{\pi} \int_0^{\pi} \mathrm{d}k \;\frac{-2\omega \kappatwo(1 - \cos k) - 8 (\kappatwo)^2 (1 - \cos k)^2}{\omega^2 + 8 \omega \kappatwo(1 - \cos k)}
        \\
        C_{ij} &= \frac{1}{\pi} \int_0^{\pi} \mathrm{d}k \;\cos(kj)\bigl[(u+v)^2\,(u^2+v^2)\bigr].
    \end{align}
\end{subequations}
The first two integrals can be analytically solved, yielding
\begin{subequations}
    \begin{align}
        \braket{\opadag \opa} &= +\frac{\kappatwo}{\omega} + \frac{1}{8}
        \Bigg(
            \sqrt{\frac{\omega}{\omega + 16\kappatwo}} - 1
        \Bigg),
        \\
        \braket{\opa^2} &= -\frac{\kappatwo}{\omega} + \frac{1}{8}
        \Bigg(
            \sqrt{\frac{\omega}{\omega + 16\kappatwo}} - 1
        \Bigg).
    \end{align}
\end{subequations}

\bibliography{references}

@article{magoni2024coherent,
  title={Coherent spin-phonon scattering in facilitated Rydberg lattices},
  author={Magoni, Matteo and Nill, Chris and Lesanovsky, Igor},
  journal={Physical Review Letters},
  volume={132},
  number={13},
  pages={133401},
  year={2024},
  publisher={APS},
  doi={10.1103/PhysRevLett.132.133401}
}

@article{magoni2021emergent,
  title={Emergent Bloch oscillations in a kinetically constrained Rydberg spin lattice},
  author={Magoni, Matteo and Mazza, Paolo P and Lesanovsky, Igor},
  journal={Physical Review Letters},
  volume={126},
  number={10},
  pages={103002},
  year={2021},
  publisher={APS},
  doi={10.1103/PhysRevLett.126.103002}
}

@book{mandel1995optical,
  title={Optical Coherence and Quantum Optics},
  author={Mandel, L and Wolf, E},
  year={1995},
  publisher={Cambridge University Press},
  doi={10.1017/CBO9781139644105},
  isbn={9780521417112},
  url={https://doi.org/10.1017/CBO9781139644105}
}

@article{hollerith2019quantum,
  title={Quantum gas microscopy of Rydberg macrodimers},
  author={Hollerith, Simon and Zeiher, Johannes and Rui, Jun and Rubio-Abadal, Antonio and Walther, Valentin and Pohl, Thomas and Stamper-Kurn, Dan M and Bloch, Immanuel and Gross, Christian},
  journal={Science},
  volume={364},
  number={6441},
  pages={664--667},
  year={2019},
  publisher={American Association for the Advancement of Science},
  doi={10.1126/science.aaw4150}
}

@article{vidal2003efficient,
  title={Efficient classical simulation of slightly entangled quantum computations},
  author={Vidal, Guifr{\'e}},
  journal={Physical Review Letters},
  volume={91},
  number={14},
  pages={147902},
  year={2003},
  publisher={APS},
  doi={10.1103/PhysRevLett.91.147902}
}

@article{de2019observation,
  title={Observation of a symmetry-protected topological phase of interacting bosons with Rydberg atoms},
  author={De L{\'e}s{\'e}leuc, Sylvain and Lienhard, Vincent and Scholl, Pascal and Barredo, Daniel and Weber, Sebastian and Lang, Nicolai and B{\"u}chler, Hans Peter and Lahaye, Thierry and Browaeys, Antoine},
  journal={Science},
  volume={365},
  number={6455},
  pages={775--780},
  year={2019},
  publisher={American Association for the Advancement of Science},
  doi={10.1126/science.aav9105}
}

@article{barredo2016atom,
  title={An atom-by-atom assembler of defect-free arbitrary two-dimensional atomic arrays},
  author={Barredo, Daniel and De L{\'e}s{\'e}leuc, Sylvain and Lienhard, Vincent and Lahaye, Thierry and Browaeys, Antoine},
  journal={Science},
  volume={354},
  number={6315},
  pages={1021--1023},
  year={2016},
  publisher={American Association for the Advancement of Science},
  doi={10.1126/science.aah3778}
}

@article{endres2016atom,
  title={Atom-by-atom assembly of defect-free one-dimensional cold atom arrays},
  author={Endres, Manuel and Bernien, Hannes and Keesling, Alexander and Levine, Harry and Anschuetz, Eric R and Krajenbrink, Alexandre and Senko, Crystal and Vuletic, Vladan and Greiner, Markus and Lukin, Mikhail D},
  journal={Science},
  volume={354},
  number={6315},
  pages={1024--1027},
  year={2016},
  publisher={American Association for the Advancement of Science},
  doi={10.1126/science.aah3752}
}

@article{barredo2018synthetic,
  title={Synthetic three-dimensional atomic structures assembled atom by atom},
  author={Barredo, Daniel and Lienhard, Vincent and De Leseleuc, Sylvain and Lahaye, Thierry and Browaeys, Antoine},
  journal={Nature},
  volume={561},
  number={7721},
  pages={79--82},
  year={2018},
  publisher={Nature Publishing Group UK London},
  doi={10.1038/s41586-018-0450-2}
}

@article{labuhn2016tunable,
  title={Tunable two-dimensional arrays of single Rydberg atoms for realizing quantum Ising models},
  author={Labuhn, Henning and Barredo, Daniel and Ravets, Sylvain and De L{\'e}s{\'e}leuc, Sylvain and Macr{\`\i}, Tommaso and Lahaye, Thierry and Browaeys, Antoine},
  journal={Nature},
  volume={534},
  number={7609},
  pages={667--670},
  year={2016},
  publisher={Nature Publishing Group UK London},
  doi={10.1038/nature18274}
}

@article{bernien2017probing,
  title={Probing many-body dynamics on a 51-atom quantum simulator},
  author={Bernien, Hannes and Schwartz, Sylvain and Keesling, Alexander and Levine, Harry and Omran, Ahmed and Pichler, Hannes and Choi, Soonwon and Zibrov, Alexander S and Endres, Manuel and Greiner, Markus and others},
  journal={Nature},
  volume={551},
  number={7682},
  pages={579--584},
  year={2017},
  publisher={Nature Publishing Group UK London},
  doi={10.1038/nature24622}
}

@article{schauss2015crystallization,
  title={Crystallization in Ising quantum magnets},
  author={Schau{\ss}, Peter and Zeiher, Johannes and Fukuhara, Takeshi and Hild, Sebastian and Cheneau, Marc and Macr{\`\i}, Tommaso and Pohl, Thomas and Bloch, Immanuel and Gro{\ss}, Christian},
  journal={Science},
  volume={347},
  number={6229},
  pages={1455--1458},
  year={2015},
  publisher={American Association for the Advancement of Science},
  doi={10.1126/science.1258351}
}

@article{lienhard2018observing,
  title={Observing the space-and time-dependent growth of correlations in dynamically tuned synthetic Ising models with antiferromagnetic interactions},
  author={Lienhard, Vincent and de L{\'e}s{\'e}leuc, Sylvain and Barredo, Daniel and Lahaye, Thierry and Browaeys, Antoine and Schuler, Michael and Henry, Louis-Paul and L{\"a}uchli, Andreas M},
  journal={Physical Review X},
  volume={8},
  number={2},
  pages={021070},
  year={2018},
  publisher={APS},
  doi={10.1103/PhysRevX.8.021070}
}

@article{guardado2018probing,
  title={Probing the quench dynamics of antiferromagnetic correlations in a 2D quantum Ising spin system},
  author={Guardado-Sanchez, Elmer and Brown, Peter T and Mitra, Debayan and Devakul, Trithep and Huse, David A and Schau{\ss}, Peter and Bakr, Waseem S},
  journal={Physical Review X},
  volume={8},
  number={2},
  pages={021069},
  year={2018},
  publisher={APS},
  doi={10.1103/PhysRevX.8.021069}
}

@article{barredo2015coherent,
  title={Coherent excitation transfer in a spin chain of three Rydberg atoms},
  author={Barredo, Daniel and Labuhn, Henning and Ravets, Sylvain and Lahaye, Thierry and Browaeys, Antoine and Adams, Charles S},
  journal={Physical Review Letters},
  volume={114},
  number={11},
  pages={113002},
  year={2015},
  publisher={APS},
  doi={10.1103/PhysRevLett.114.113002}
}

@article{lukin2001dipole,
  title={Dipole blockade and quantum information processing in mesoscopic atomic ensembles},
  author={Lukin, Mikhail D and Fleischhauer, Michael and Cote, Robin and Duan, LuMing and Jaksch, Dieter and Cirac, J Ignacio and Zoller, Peter},
  journal={Physical review letters},
  volume={87},
  number={3},
  pages={037901},
  year={2001},
  publisher={APS}
}

@article{weimer2010rydberg,
  title={A Rydberg quantum simulator},
  author={Weimer, Hendrik and M{\"u}ller, Markus and Lesanovsky, Igor and Zoller, Peter and B{\"u}chler, Hans Peter},
  journal={Nature Physics},
  volume={6},
  number={5},
  pages={382--388},
  year={2010},
  publisher={Nature Publishing Group UK London},
  doi={10.1038/nphys1614}
}

@article{ates2007antiblockade,
  title={Antiblockade in Rydberg excitation of an ultracold lattice gas},
  author={Ates, Cenap and Pohl, Thomas and Pattard, Thomas and Rost, Jan M},
  journal={Physical Review Letters},
  volume={98},
  number={2},
  pages={023002},
  year={2007},
  publisher={APS},
  doi={10.1103/PhysRevLett.98.023002}
}

@book{gallagher1994rydberg,
  title={Rydberg Atoms},
  author={Gallagher, Thomas F.},
  year={1994},
  publisher={Cambridge University Press},
  doi={10.1017/CBO9780511524530}
}

@article{browaeys2020many,
  title={Many-body physics with individually controlled Rydberg atoms},
  author={Browaeys, Antoine and Lahaye, Thierry},
  journal={Nature Physics},
  volume={16},
  number={2},
  pages={132--142},
  year={2020},
  publisher={Nature Publishing Group UK London},
  doi={10.1038/s41567-019-0733-z}
}

@article{bloch1929quantenmechanik,
  title={{\"U}ber die quantenmechanik der elektronen in kristallgittern},
  author={Bloch, Felix},
  journal={Zeitschrift f{\"u}r Physik},
  volume={52},
  number={7},
  pages={555--600},
  year={1929},
  publisher={Springer},
  doi={10.1007/BF01339455}
}

@article{srednicki1994chaos,
  title={Chaos and quantum thermalization},
  author={Srednicki, Mark},
  journal={Physical Review E},
  volume={50},
  number={2},
  pages={888},
  year={1994},
  publisher={APS},
  doi={10.1103/PhysRevE.50.888}
}

@article{deutsch1991quantum,
  title={Quantum statistical mechanics in a closed system},
  author={Deutsch, Josh M},
  journal={Physical Review A},
  volume={43},
  number={4},
  pages={2046},
  year={1991},
  publisher={APS},
  doi={10.1103/PhysRevA.43.2046}
}

@Article{samajdar2018numerical,
  title={Numerical study of the chiral Z 3 quantum phase transition in one spatial dimension},
  author={Samajdar, Rhine and Choi, Soonwon and Pichler, Hannes and Lukin, Mikhail D and Sachdev, Subir},
  journal={Physical Review A},
  volume={98},
  number={2},
  pages={023614},
  year={2018},
  publisher={APS},
  doi={10.1103/PhysRevA.98.023614}
}

@Article{Semeghini2021,
  author   = {G. Semeghini and H. Levine and A. Keesling and S. Ebadi and T. T. Wang and D. Bluvstein and R. Verresen and H. Pichler and M. Kalinowski and R. Samajdar and A. Omran and S. Sachdev and A. Vishwanath and M. Greiner and V. Vuleti{\'{c}} and M. D. Lukin},
  journal  = {Science},
  title    = {Probing topological spin liquids on a programmable quantum simulator},
  year     = {2021},
  number   = {6572},
  pages    = {1242-1247},
  volume   = {374},
  doi      = {10.1126/science.abi8794},
}

@article{li2013nonadiabatic,
  title={Nonadiabatic Motional Effects and Dissipative Blockade for Rydberg Atoms Excited<? format?> from Optical Lattices or Microtraps},
  author={Li, W and Ates, C and Lesanovsky, I},
  journal={Physical Review Letters},
  volume={110},
  number={21},
  pages={213005},
  year={2013},
  publisher={APS},
  doi={10.1103/PhysRevLett.110.213005}
}

@article{schachenmayer2010dynamical,
  title={Dynamical crystal creation with polar molecules or Rydberg atoms in optical lattices},
  author={Schachenmayer, J and Lesanovsky, I and Micheli, A and Daley, AJ},
  journal={New Journal of Physics},
  volume={12},
  number={10},
  pages={103044},
  year={2010},
  publisher={IOP Publishing},
  doi={10.1088/1367-2630/12/10/103044}
}

@article{ohler2023quantum,
  title={Quantum spin liquids of Rydberg excitations in a honeycomb lattice induced by density-dependent Peierls phases},
  author={Ohler, Simon and Kiefer-Emmanouilidis, Maximilian and Fleischhauer, Michael},
  journal={Physical Review Research},
  volume={5},
  number={1},
  pages={013157},
  year={2023},
  publisher={APS},
  doi={10.1103/PhysRevResearch.5.013157}
}

@article{weimer2008quantum,
  title = {Quantum Critical Behavior in Strongly Interacting Rydberg Gases},
  author = {Weimer, Hendrik and L\"ow, Robert and Pfau, Tilman and B\"uchler, Hans Peter},
  journal = {Phys. Rev. Lett.},
  volume = {101},
  issue = {25},
  pages = {250601},
  numpages = {4},
  year = {2008},
  month = {Dec},
  publisher = {American Physical Society},
  doi = {10.1103/PhysRevLett.101.250601},
  url = {https://link.aps.org/doi/10.1103/PhysRevLett.101.250601}
}

@book{britanak2007dct,
title = {Discrete Cosine and Sine Transforms},
publisher = {Academic Press},
address = {Oxford},
year = {2007},
isbn = {978-0-12-373624-6},
author = {Vladimir Britanak and Patrick C. Yip and K.R. Rao}
}

@article{Croy2011,
journal={The European Physical Journal B: Condensed Matter and Complex Systems},
author={A. Croy and P. Cain and M. Schreiber},
title={Anderson localization in 1D systems with correlated disorder},
year={2011},
month={July},
pages={107-112},
volume={82},
number={2},
doi={10.1140/epjb/e2011-20212-1},
url={https://ideas.repec.org/a/spr/eurphb/v82y2011i2p107-112.html},
}

@article{Makse1996,
  title = {Method for generating long-range correlations for large systems},
  author = {Makse, Hern\'an A. and Havlin, Shlomo and Schwartz, Moshe and Stanley, H. Eugene},
  journal = {Phys. Rev. E},
  volume = {53},
  issue = {5},
  pages = {5445--5449},
  numpages = {0},
  year = {1996},
  month = {May},
  publisher = {American Physical Society},
  doi = {10.1103/PhysRevE.53.5445},
  url = {https://link.aps.org/doi/10.1103/PhysRevE.53.5445}
}

@misc{wolfram_2026_elliptice, 
author={{Wolfram Research}}, 
title="{EllipticE}", 
year="2022", 
howpublished="\url{https://reference.wolfram.com/language/ref/EllipticE.html}", 
note=[Accessed:09-July-2026]
}

@article{Kramer1993,
doi = {10.1088/0034-4885/56/12/001},
url = {https://doi.org/10.1088/0034-4885/56/12/001},
year = {1993},
month = {dec},
publisher = {},
volume = {56},
number = {12},
pages = {1469},
author = {B Kramer and A MacKinnon},
title = {Localization: theory and experiment},
journal = {Reports on Progress in Physics}
}

@article{Anderson1958,
  title = {Absence of Diffusion in Certain Random Lattices},
  author = {Anderson, P. W.},
  journal = {Phys. Rev.},
  volume = {109},
  issue = {5},
  pages = {1492--1505},
  numpages = {0},
  year = {1958},
  month = {Mar},
  publisher = {American Physical Society},
  doi = {10.1103/PhysRev.109.1492},
  url = {https://link.aps.org/doi/10.1103/PhysRev.109.1492}
}

@article{jaksch2000fast,
  title={Fast quantum gates for neutral atoms},
  author={Jaksch, Dieter and Cirac, Juan Ignacio and Zoller, Peter and Rolston, Steve L and C{\^o}t{\'e}, Robin and Lukin, Mikhail D},
  journal={Physical Review Letters},
  volume={85},
  number={10},
  pages={2208},
  year={2000},
  publisher={APS}
}

\end{document}